\documentclass[journal]{IEEEtran}
\usepackage{tipa}
\usepackage{bbm}
\usepackage{mathrsfs}
\usepackage{cite,url,subfigure,epsfig,graphicx}
\usepackage{amssymb,amsmath}
\usepackage{indentfirst}
\usepackage{algorithmic}
\usepackage{algorithm}
\usepackage{pdfpages}
\usepackage{times,verbatim,amsfonts,amsmath,color}
\usepackage{pifont}
\usepackage{multirow}
\usepackage{booktabs}
\usepackage{adjustbox}
\usepackage{hyperref}

\IEEEoverridecommandlockouts
\makeatletter
\def\ps@headings{\def\@oddhead{\mbox{}\scriptsize\rightmark \hfil \thepage}\def\@evenhead{\scriptsize\thepage \hfil \leftmark\mbox{}}\def\@oddfoot{}\def\@evenfoot{}}
\makeatother 

\newcommand{\tabincell}[2]{\begin{tabular}{@{}#1@{}}#2\end{tabular}}

\usepackage{geometry}
\begin{document}
\title{A Survey on ChatGPT: AI–Generated Contents, Challenges, and Solutions}
\author{
\IEEEauthorblockN{Yuntao~Wang, Yanghe~Pan, Miao~Yan, Zhou~Su\IEEEauthorrefmark{1}, and Tom H. Luan}\\
\IEEEauthorblockA{
School of Cyber Science and Engineering, Xi'an Jiaotong University, Xi'an, China\\
\IEEEauthorrefmark{1}Corresponding Author: zhousu@ieee.org
}}

\author{Yuntao~Wang, Yanghe~Pan, Miao~Yan, Zhou~Su, and Tom H. Luan
\thanks{Manuscript accepted July 27, 2023 by IEEE Open Journal of the Computer Society with DOI:10.1109/OJCS.2023.3300321. This work was supported in part by NSFC (nos. U22A2029, U20A20175), and the Fundamental Research Funds for the Central Universities. (\emph{Corresponding author: Zhou~Su}).}
\thanks{Y. Wang, Y. Pan, M. Yan, Z. Su, and T. H. Luan are with the School of Cyber Science and Engineering, Xi'an Jiaotong University, Xi'an, China (e-mail: zhousu@ieee.org).}
\thanks{\textbf{Cite as:} Y. Wang, Y. Pan, M. Yan, Z. Su, and T. H. Luan, ``A Survey on ChatGPT: AI-Generated Contents, Challenges, and Solutions", \emph{IEEE Open Journal of the Computer Society}, early access, 2023, \textbf{DOI:}10.1109/OJCS.2023.3300321.}}
\maketitle

\begin{abstract}
With the widespread use of large artificial intelligence (AI) models such as ChatGPT, AI-generated content (AIGC) has garnered increasing attention and is leading a paradigm shift in content creation and knowledge representation.
AIGC uses generative large AI algorithms to assist or replace humans in creating massive, high-quality, and human-like content at a faster pace and lower cost, based on user-provided prompts. Despite the recent significant progress in AIGC, security, privacy, ethical, and legal challenges still need to be addressed.
This paper presents an in-depth survey of working principles, security and privacy threats, state-of-the-art solutions, and future challenges of the AIGC paradigm. Specifically, we first explore the enabling technologies, general architecture of AIGC, and discuss its working modes and key characteristics. Then, we investigate the taxonomy of security and privacy threats to AIGC and highlight the ethical and societal implications of GPT and AIGC technologies. Furthermore, we review the state-of-the-art AIGC watermarking approaches for regulatable AIGC paradigms regarding the AIGC model and its produced content. Finally, we identify future challenges and open research directions related to AIGC.
\end{abstract}


\IEEEpeerreviewmaketitle
\section{Introduction}
Artificial intelligence-generated content (AIGC) refers to the use of generative AI algorithms to assist or replace humans in creating rich personalized and high-quality content at a faster pace and lower cost, based on user inputs or requirements \cite{wu2023ai,xu2023unleashing,cao2023comprehensive}.
AIGC encompasses a vast range of synthetic content, including text (e.g., poems), images (e.g., artwork), audio (e.g., music), video (e.g., cartoons), augmented training samples, and interactive 3D content (e.g., virtual avatars, assets, and environments).
As a complement to traditional content creation paradigms such as professional generated content (PGC) and user generated content (UGC), the promising AIGC paradigm allows the low-cost production of massive amounts of contents in an automated and efficient manner \cite{du2023enabling}, which is beneficial to various emerging applications such as metaverse \cite{9880528} and digital twin \cite{10090432}.
For instance, in Roblox (an interactive metaverse game), AIGC can produce personalized skins and 3D game scenarios for avatars, allowing users to play, collaborate, and socialize in an immersive virtual space. According to Gartner \cite{GartnerRep}, generative AI algorithms are projected to produce around 10\% of all data by 2025.

\begin{figure}[!t]
\centering \setlength{\abovecaptionskip}{-0.025cm}
  \includegraphics[width=6cm]{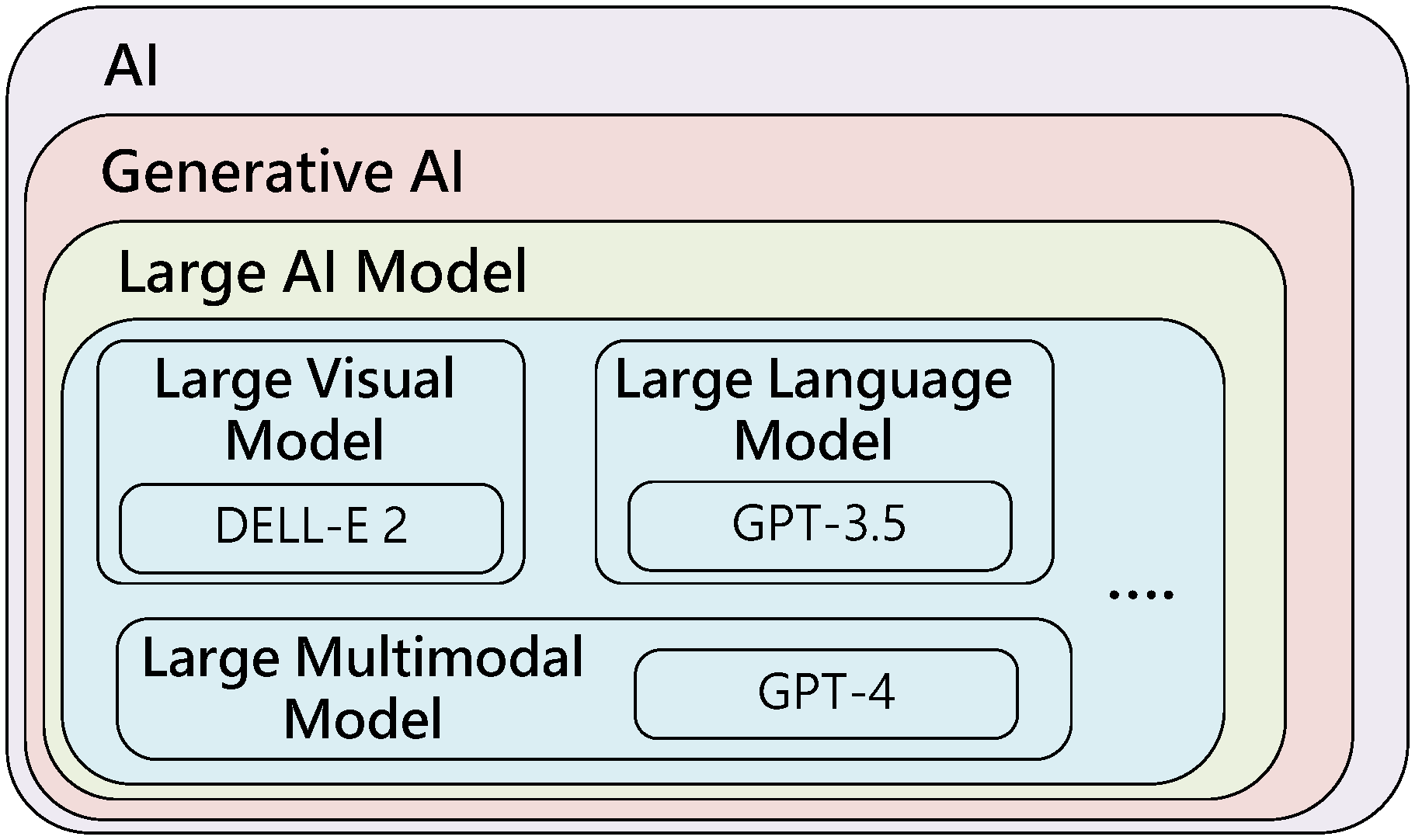}
  \caption{Relation between existing representative large AI models and AIGC. {Generative AI algorithms are a class of AI algorithms that create new content in various forms (e.g., images, text, and music) by learning underlying patterns from training data. AIGC encompasses a broader scope and includes not only generative AI algorithms but also other AI techniques such as natural language processing and computer vision. A large AI model refers to any neural network architecture that has large number of parameters, such as large visual model (LVM), large language model (LLM), and large multimodal
model.}}\label{fig:Relation}\vspace{-2mm}
\end{figure}

Technically, AIGC typically consists of two phases \cite{cao2023comprehensive}: (i) extracting and understanding user intent information, and (ii) producing desired content based on the extracted intentions.
In November 2022, OpenAI released ChatGPT, a versatile language model that can generate code, write stories, perform machine translation, conduct semantic analysis, and etc. By January 2023, nearly 13 million users were interacting with ChatGPT daily \cite{ChatGPTUser}.
ChatGPT is a variant of the generative pre-training transformer (GPT), a transformer-based large language model (LLM) that can understand human languages and create human-like text (e.g., stories and articles) \cite{10113601}, as shown in Fig.~\ref{fig:Relation}. With recent advancements in LLMs such as ChatGPT and its successor GPT-4 {(which is a large multimodal model)}, the capabilities of AIGC have been significantly strengthened to perform more complex tasks (e.g., multimodal tasks) with higher accuracy, thanks to the better intent extraction offered by LLMs \cite{yang2023harnessing}.
Driven by technological advances and increasing demand, AIGC has gained worldwide attention and shown great potential for various applications, including entertainment, advertising, art, and education. Tech giants including OpenAI, Google, Microsoft, NVIDIA, and Baidu have announced their adventure to AIGC and developed their own AIGC products.

In the AIGC era, larger datasets are the ``fuel", larger foundation model serves as the ``engine", and extensive computing power acts as the ``accelerator".
For ChatGPT fine-tuned from the GPT-3.5 model, its training dataset consists of nearly 1 trillion words, approximately 45TB in size \cite{gpt3everything}, and it integrates several AI technologies such as self-supervised learning, reinforcement learning, and prompt learning in pre-training GPT. The computing power required for ChatGPT training is around 3640 PetaFLOPs per day, equivalent to computing 10 quadrillion times per second and would take 3640 days to complete \cite{ChatGPTComputing}.
Under the engineering combination of big data, big models, and big computing power, ChatGPT demonstrates powerful emergent abilities in learning new features and higher-level patterns and automating the creation of valuable contents based on users' multimodal prompts.
Apart from the benefits brought by massive training data and extensive computational power, ChatGPT integrates a series of new technologies.
For instance, ChatGPT employs chain-of-thought (CoT) prompting \cite{wei2023chainofthought}, which enables the pretrained LLM to explain its reasoning via step-by-step reasoning in few-shot and zero-shot learning settings. In addition, reinforcement learning from human feedback (RLHF) \cite{NEURIPS2022b1efde53} is integrated to help ChatGPT better understand human preferences by training a reward model which incorporates human feedbacks and fine-tuning the LLM with reinforcement learning. 
Moreover, in computer vision (CV), {large visual
models (LVMs) such as} stable diffusion \cite{rombach2022high} (developed by the start-up Stability AI) and DALL-E 2 \cite{Marcus2022AVP} (developed by OpenAI) in 2022 have demonstrated enormous success to produce high-resolution and natural-looking images from complex and diverse textual descriptions.


\subsection{Motivation}\label{sec:Challenges}
Despite the bright future of AIGC, security and privacy concerns pose significant obstacles to its widespread adoption. Throughout the life-cycle of AIGC services, a number of security flaws, privacy breaches, trust concerns, and ethical issues could arise from the pervasive data collection, smart model/data thefts, to the distribution of massive phishing emails.

\emph{1) Security vulnerabilities.}
AIGC models face security threats at every stage of the lifecycle. For instance, during model training, attackers may use poisoned or adversarial samples to degrade model performance \cite{tian2022comprehensive}, or launch backdoor attacks to manipulate model outcomes \cite{chen2017targeted}; after model deployment, attackers may steal the AIGC model or its partial functioning via smart model theft attacks \cite{krishna2019thieves}.
As large AIGC models such as ChatGPT adopt more complex strategies than general-purpose models, more security threats (e.g., jailbreak \cite{li2023multi} and prompt injection \cite{perez2022ignore}) that can be brand new may emerge in AIGC. Additionally, generative AI models still face technical limitations regarding transparency, robustness, and bias/ discrimination. 

\emph{2) Privacy violations.}
The success of AIGC models relies heavily on extensive training datasets that may inevitably contain users' sensitive and private information. For instance, ChatGPT is able to memorize conversation-related items as well as user inputs, cookies, and logs when interacting with users \cite{ChatGPToutage1}. This introduces new possibilities for data misuse and criminal activity in AIGC. According to a recent study \cite{carlini2021extracting} on the black-box GPT-2 model, adversaries can use prompt injections and public text features to recover up to 67\% of the training text, including personal names, addresses, and phone numbers, from the AI memory. In March 2023, Italy banned the use of ChatGPT due to privacy compliance concerns \cite{ChatGPTBanItaly}. 

\emph{3) Trust issues.}
The rapid development of AIGC technology has made it increasingly easier to create and spread disinformation and false evidence such as deep fake contents and fake news \cite{chatgptfakenews}. This has led to the emergence of new types of criminal activity, such as AI fraud, defamation, identity theft, and impersonation \cite{weidinger2021ethical}. For instance, ChatGPT can produce misleading and unethical responses, and individuals with malicious intent can take advantage of its capacity to compose flawless text for fraud, replicate speech patterns for impersonation, and develop malicious codes for hacking. It raises huge needs for to establish traceable provenance and regulations for materials produced by generative AI models to ensure accountability. 

\emph{4) Ethical effects.}
As a double-edged sword, AIGC technologies also have detrimental effects on human societies and can be abused to distribute malware, ransomware, and phishing mails. For instance, ChatGPT's ability to produce instant and convincing dialogues can make it easier to craft phishing emails that trick recipients into clicking on harmful links, downloading malware, or revealing confidential information \cite{scam}. Additionally, AIGC can facilitate cheating in classes, plagiarism in the arts, and academic paper fraud, making such misbehaviors easier to commit and harder to detect.


\subsection{Comparison with Existing Surveys and Contributions of Our Work}\label{subsec:Contributions}
As a hot topic, a significant number of research efforts have investigated the potential of AIGC from various perspectives. The notable survey papers in this area are as below. 
Wu \emph{et al}. \cite{wu2023ai} discuss key concepts, cutting-edge capabilities, industrial chains, generation modes, and critical challenges of AIGC.
Cao \emph{et al}. \cite{cao2023comprehensive} present the history of generative AI models and the role of ChatGPT in the AIGC era. Besides, the authors in \cite{cao2023comprehensive} systematically review the basic concepts and recent advances in AIGC from both unimodal and multimodal interactions.
Zhang \emph{et al}. \cite{zhang2023one} survey the underlying technologies, applications, and challenges of ChatGPT, as well as its utilization in achieving general-purpose AIGC.
Ray \cite{ray2023chatgpt} comprehensively survey the development history, underlying technologies, applications, key challenges, and future directions of ChatGPT.
Gozalo-Brizuela \emph{et al}. \cite{gozalo2023chatgpt} offer a taxonomy of the generative AI models including text-to-image, text-to-3D, image-to-text, text-to-video, text-to-audio, text-to-text, text-to-code, text-to-science models.
Zhang \emph{et al}. \cite{zhang2023survey} survey the recent progress and mechanisms of graph diffusion models, and discuss their role in three scientific AIGC applications: molecule, protein, and material.
Yang \emph{et al}. \cite{yang2023harnessing} comprehensively review the utilization of LLMs from the perspectives of data, models, and downstream missions.
Zhou \emph{et al}. \cite{zhou2023comprehensive} provide an in-depth survey on recent advancements, key challenges, and opportunities for pretrained large models in text, image, graph, and other modalities.
Zhang \emph{et al}. \cite{zhang2023complete} offer an in-depth review of the fundamental technologies behind AIGC, industrial applications, and technical development in AIGC associated with text, image, and other more complex tasks.
Xu \emph{et al}. \cite{xu2023unleashing} investigate mobile AIGC networks by deploying AIGC applications at mobile edge networks to deliver efficient, personalized, and privacy-preserving AIGC services.
The authors in \cite{xu2023unleashing} also comprehensively survey the cooperative cloud-edge-mobile technologies to support efficient mobile AIGC services and discuss potential applications and use cases in mobile AIGC networks.

\begin{table}[!t]
   \centering \setlength{\abovecaptionskip}{0cm}
    \caption{A Comparison of Our Work with Relevant Surveys}\label{contribution}
    \resizebox{1.03\linewidth}{!}{
        \begin{tabular}{|c|c|l|}
        \hline
        \textbf{Year.} &\textbf{Refs.} &\textbf{Contribution} \\ \hline 
        {2023} &\cite{wu2023ai} &\tabincell{l}{Discussions on key concepts, cutting-edge capabilities, industrial chain, \\generation modes, and critical challenges of AIGC.} \\ \hline

        {2023} &\cite{cao2023comprehensive} &\tabincell{l}{Study on the history of generative AI and the role of ChatGPT in AIGC, \\and review on basic concepts and recent advances in AIGC from \\unimodal \& multimodal interactions.} \\ \hline

        {2023} &\cite{zhang2023one} &\tabincell{l}{Review on underlying technologies, applications, and challenges of ChatGPT, \\as well as the utilization of ChatGPT to realize general-purpose AIGC.} \\ \hline

        {2023} &\cite{ray2023chatgpt} &\tabincell{l}{Review on development history, underlying technologies, applications, \\key challenges, and future directions of ChatGPT.} \\ \hline

        {2023} &\cite{gozalo2023chatgpt} &\tabincell{l}{A taxonomy of generative AI models in terms of text-to-image, image-to-text, \\text-to-3D, text-to-video, text-to-audio, text-to-text, text-to-code, text-to-science.} \\ \hline

        {2023} &\cite{zhang2023survey} &\tabincell{l}{Survey on recent progress and mechanisms of graph diffusion models and its \\role in three scientific AIGC applications, i.e., molecule, protein, and material.} \\ \hline

        {2023} &\cite{yang2023harnessing}  &\tabincell{l}{Review on utilization of LLMs from the perspectives of data, models, and \\downstream missions.} \\ \hline

        {2023} &\cite{zhou2023comprehensive}  &\tabincell{l}{Survey on recent advancements, key challenges, and opportunities for \\pretrained large models in text, image, graph, and other modalities.} \\ \hline

        {2023} &\cite{zhang2023complete} &\tabincell{l}{An in-depth review on fundamental technologies behind AIGC, industrial \\applications, and technical development of AIGC tasks from text, image \\and other more complex tasks.} \\ \hline

        {2023} &\cite{xu2023unleashing} &\tabincell{l}{A comprehensive survey on mobile AIGC networks, including cooperative\\ cloud-edge-mobile technologies to support efficient mobile AIGC services and \\potential applications in mobile AIGC networks.} \\ \hline

        {Now} &{Ours} &\tabincell{l}{A comprehensive survey of the general architecture, working modes, key \\characteristics, and applications of AIGC, discussions on the security and \\privacy threats to AIGC, IP protection for AIGC models and contents, \\state-of-the-art solutions to regulatable AIGC paradigms, and open directions.} \\ \hline
        \end{tabular}}
\end{table}

Unlike the aforementioned existing surveys on AIGC, the goal of this survey is to thoroughly discuss the fundamentals, security/privacy/ethical issues, and regulations of AIGC, including the general architecture, enabling technologies, working modes, key characteristics, applications, prototypes, existing/potential security and privacy threats to AIGC, intellectual property (IP) protection for AIGC models and contents, state-of-the-art solutions to regulatable AIGC paradigms, and open directions in AIGC. Table~\ref{contribution} compares the contributions of our survey and previous related surveys in the field of AIGC.
The main contributions of this work are four-fold.
\begin{itemize}
  \item We investigate the working principles of the AIGC paradigm, including the three-layer general architecture, enabling technologies (i.e., generative AI algorithms, pretrained large AI models, multimodal technology, and RLHF), working modes (i.e., assisted and autonomous content creation), key characteristics, applications, and modern prototypes. The paradigm shift brought by AIGC in content creation as well as knowledge representation and usage are also discussed. 
  \item We comprehensively survey the existing/potential security and privacy threats in the AIGC paradigm from four perspectives: security, privacy, trust, and ethics. Besides, the advanced defense mechanisms in both academia and industry are examined and their feasibility in AIGC is discussed. 
  \item We discuss the IP protection concerns in AIGC and the related crucial challenges. The state-of-the-art watermark-based IP protection solutions for both AIGC models and contents are systematically reviewed, as well as the threats and countermeasures to AIGC watermarking.
  \item Lastly, we discuss open research issues and point out future research directions from the perspectives of green AIGC architecture, explainable AIGC models, distributed and scalable AIGC algorithms, trustworthy and regulatable AIGC services, and secure-by-design AIGC paradigm, toward building the most efficient and secure AIGC system.
\end{itemize}

\subsection{Paper Organization}\label{subsec:organization}
\begin{figure}[!t]
\centering \setlength{\abovecaptionskip}{-0.1cm}
\includegraphics[width=7.8cm,height=14.cm]{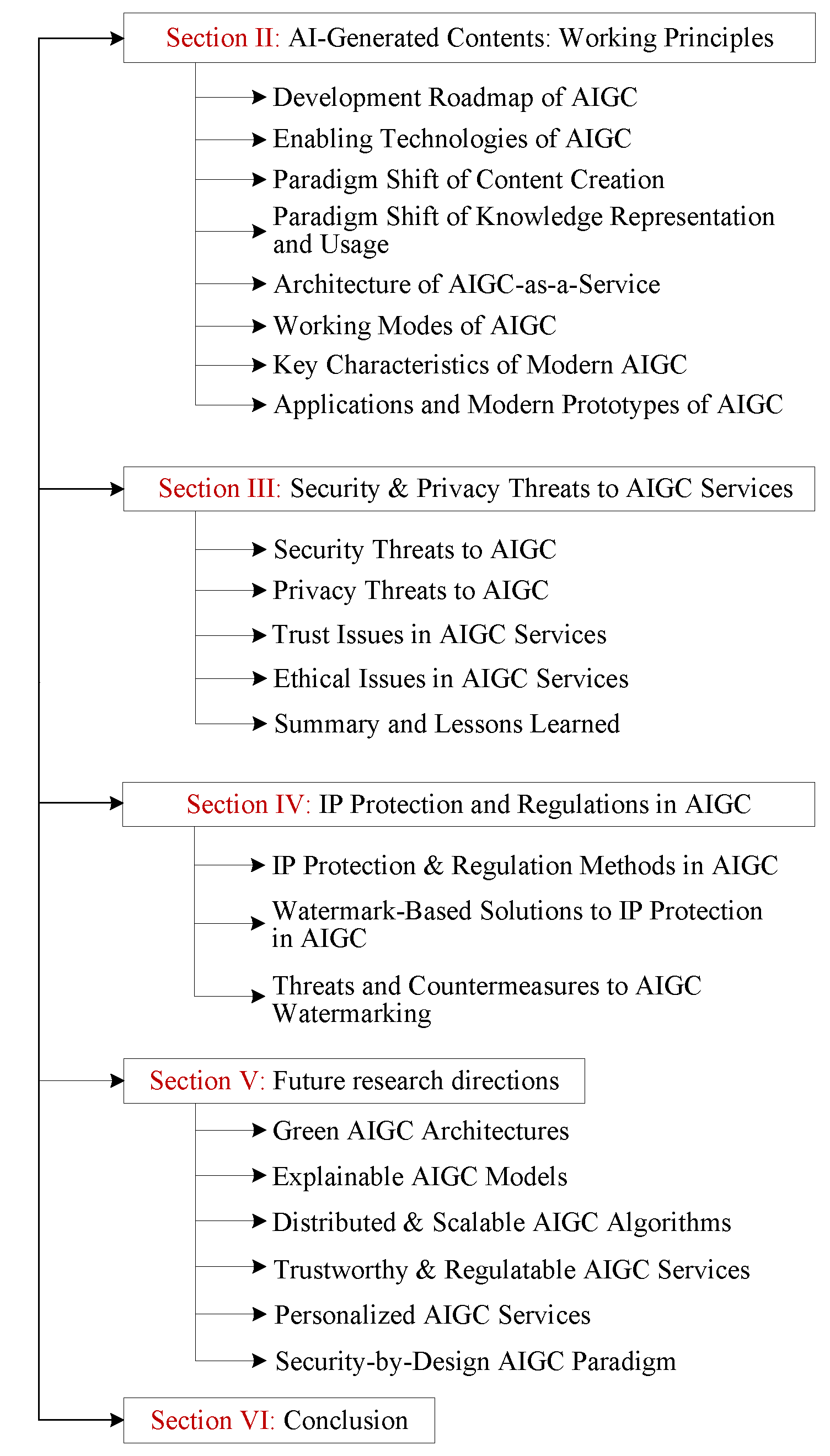}
  \caption{Organization structure of this survey paper.}\label{fig:organization}
\end{figure}

The remainder of this paper is organized as below.  In Section~\ref{sec:OVERVIEW}, we introduce the working principles of AIGC. Section~\ref{sec:Threat} discusses the taxonomy of security and privacy concerns in AIGC, along with state-of-the-art countermeasures. Section~\ref{sec:Regu} presents IP protection and regulation of AIGC models and contents. Section~\ref{sec:FUTUREWORK} explores the future research directions. Finally, Section~\ref{sec:CONSLUSION} draws the conclusions. The organization structure of this survey is illustrated in Fig.~\ref{fig:organization}.

\section{AI-Generated Contents: Working Principles}\label{sec:OVERVIEW}
In this section, we first introduce the development roadmap and enabling technologies of AIGC. Then, we discuss the paradigm shift in content creation paradigms as well as knowledge representation and usage paradigms. After that, we present the general architecture, working modes, key characteristics, applications, and modern prototypes of the AIGC.

\subsection{Development Roadmap of AIGC}\label{subsec:History}
\begin{itemize}
  \item \emph{Early Stage (before mid 2010s):} In 1957, Hiller and Isaacson invented the first computer-synthesized music:  \emph{Iliac Suite2}. Despite this early success, due to the high costs and difficulties in commercialization, AIGC did not achieve much success in the 20th century. With the advancement of deep learning and increasing computing power, early attempts at AIGC have resurfaced. For example, in 2007, the first AI-produced novel: \emph{On the Road} was released. Additionally, in 2012, Microsoft introduced a fully automatic simultaneous interpretation system based on deep neural networks (DNN) for automatic speech recognition and translation from Chinese to English.

  \item \emph{Development Stage (mid 2010s$\sim$2022):} In this phase, generative AI algorithms emerge and grow rapidly, laying the foundation for large model training. In 2014, Goodfellow \emph{et al.} proposed the generative adversarial network (GAN) for producing images from existing data \cite{NIPS20145ca3e9b1}. In 2017, Microsoft's AI bot \emph{Xiaoice} created the world's first poetry collection, entitled \emph{Sunshine Lost the Glass Window}. In 2017, Vaswani \emph{et al.} proposed the transformer model for NLP tasks with parallel training capacities \cite{NIPS20173f5ee243}. In 2019, DeepMind released the \emph{DVD-GAN} model, which is capable of generating continuous video \cite{clark2019adversarial}. In 2021, OpenAI launched DALL-E, which supports image generation from texts \cite{Marcus2022AVP}.

  \item \emph{Flourishing Stage (2022$\sim$now):} In this phase, with the availability of pretrained large AI models, the capabilities of AIGC have been greatly enhanced, making it practical for large-scale uses. For instance, in 2022, OpenAI launched ChatGPT as a do-it-all LMM based on based on the generative pretrained transformer (GPT), which can perform various complex tasks such as generating human-like responses and marketing copywriting \cite{ray2023chatgpt}. The release of multimodal GPT-4 in 2023 further expanded the power of large AI models. Other notable models include PaLM 2 (34B), StableLM (175B), LLaMa (65B), ERNIE-ViLG (10B), Imagen, and DALL-E 2. As these models continue to evolve, it is anticipated to see a wide range of applications and uses for AIGC technology in the coming years.
\end{itemize}

\subsection{Enabling Technologies of AIGC}\label{subsec:Technology}
The convergence of cutting-edge AI technologies including generative AI algorithms, pretrained large models, and multimodality techniques leads to the flourishing of the AIGC paradigm. 
\begin{itemize}
  \item \emph{Generative AI Algorithms.} Generative AI algorithms are a category of AI algorithms that learn patterns from large sets of data and then produce new content \cite{gozalo2023chatgpt}. With recent advances in generative AI algorithms, such as the transformer \cite{NIPS20173f5ee243} and diffusion models \cite{du2023enabling}, AI models can learn to generate realistic images, videos, and text based on patterns and data in training samples without human intervention. For instance, the popular transformer model uses a self-attention mechanism to assign different weights based on the importance of each part of the input data \cite{NIPS20173f5ee243}. Various pre-training models such as BERT and ChatGPT are based on the transformer. The diffusion model defines a Markov chain of diffusion steps, which gradually adds random noise to the data and then learn the inverse diffusion process to construct necessary data samples from the noise \cite{du2023enabling}. Stable Diffusion is a latent text-to-image diffusion model capable of producing photo-realistic images from textual descriptions \cite{rombach2022high}.

  \item \emph{Pretrained Large AI Models.} A large AI model refers to any neural network architecture that has a large number of parameters, capable of providing better representation learning and achieving higher accuracy on various tasks \cite{zhou2023comprehensive}. Essentially, flexible, large-scale and high-quality AIGC services are made possible by pretrained large models such as large visual model (LVM) and LLM. For instance, ChatGPT can process and learn from vast amounts of text data to produce coherent and naturalistic text outputs such as articles, stories, and even poetry. Besides, ChatGPT integrates a series of new technologies such as CoT prompt \cite{wei2023chainofthought} and RLHF \cite{NEURIPS2022b1efde53} to offer more interactive and personalized conversations.

  \item \emph{Multimodal Technology.} Multimodal technology empowers pretrained large models with multimodal representation capabilities, thereby enabling efficient fusion of texts, images, languages, and other modalities. As such, it greatly enriches the diversity of contents produced or processed by AI. For instance, OpenAI designed contrastive language-image pre-training (CLIP) \cite{rombach2022high}, which is capable of performing NLP understanding and CV analysis simultaneously for text-image association. CLIP uses over 4 billion text-image training data from the Internet \cite{gpt3everything}. With the support of multimodal technology, the pretrained large model has developed from the early single NLP or CV model to the multi-modal and cross-modal AIGC models.

\end{itemize}

\subsection{Paradigm Shift of Content Creation}\label{subsec:Paradigm1}
The paradigm of AIGC is revolutionizing the landscape of content generation with markable advantages such as large-scale production, high-quality output, and low unit costs, as summarized in Table~\ref{Content}.
With the use of pretrained large models (e.g., ChatGPT) and multimodal technology (e.g., CLIP), the AIGC paradigm has emerged as a powerful tool for generating diverse content types, such as articles, stories, and poetry, with a level of efficiency and affordability that was not possible before.

\begin{itemize}
  \item The invention of the Internet brought about the possibility of information dissemination and replication at zero cost. In the era of the \emph{Web 1.0 (PC Internet stage)}, the creation of high-quality content primarily fell under the professionally-generated content (PGC) paradigm \cite{9880528}, where professional organizations or individuals are the main source of content production. However, due to the limited scale of content creation, this paradigm restricted the availability and variety of content for users.

  \item With the advent of the \emph{Web 2.0 (mobile Internet stage)}, users emerged as key players in content creation, leading to the development of the user-generated content (UGC) paradigm. In complement to the PGC paradigm, users under the UGC paradigm can freely produce content on diverse platforms and share them with each other \cite{9880528}. Representatives UGC platforms include Facebook, Sina Weibo, TikTok, and others.

  \item The metaverse is recognized as the future form of the \emph{Web 3.0} \cite{9880528}, and AIGC is the key to building the metaverse system by concurrently producing massive 3D and immersive content (e.g., personalized metaverse scenes) in real time and at a low cost. With the aid of computing power and large AI models, AIGC can automatically generate or assist humans to create a large volume of content in a short amount of time. While traditional PGC and UGC paradigms face limitations regarding scale, quality, and cost, AIGC can effectively address the shortcomings of these models. 
\end{itemize}

\begin{table}[!t]
   \centering \setlength{\abovecaptionskip}{0cm}
    \caption{A Comparison of Content Creation Modes}\label{Content}
\resizebox{1.01\linewidth}{!}{
\begin{tabular}{|c|c|c|c|c|}
\hline
\textbf{} & \textbf{\begin{tabular}[c]{@{}c@{}}Content Creation\\ Mode\end{tabular}} & \textbf{Features}                                                                       & \textbf{Limitations}                                                          & \textbf{Representatives}                                  \\ \hline
Web 1.0   & PGC                                                                      & \begin{tabular}[c]{@{}c@{}}High quality\\ Low diversity\end{tabular}                    & \begin{tabular}[c]{@{}c@{}}Limited by\\ capacity of production\end{tabular}   & Web portal                                                \\ \hline
Web 2.0   & UGC                                                                      & \begin{tabular}[c]{@{}c@{}}High diversity\\ Medium cost\end{tabular}                    & \begin{tabular}[c]{@{}c@{}}Limited by \\ content quality\end{tabular}         & \begin{tabular}[c]{@{}c@{}}Facebook\\ TikTok\end{tabular} \\ \hline
Web 3.0   & AIGC                                                                     & \begin{tabular}[c]{@{}c@{}}High efficiency \\Near zero marginal cost\end{tabular} & \begin{tabular}[c]{@{}c@{}}Limited by AIGC\\ technology maturity\end{tabular} & Metaverse                                                 \\ \hline
\end{tabular} }
\end{table}

\begin{table}[!t]
   \centering \setlength{\abovecaptionskip}{0cm}
    \caption{A Comparison of Knowledge Representation and Usage Modes}\label{knowledge}
\resizebox{1.01\linewidth}{!}{
\begin{tabular}{|c|c|c|c|c|}
\hline
\textbf{\begin{tabular}[c]{@{}c@{}}Knowledge \\Representation Mode\end{tabular}} & \textbf{\begin{tabular}[c]{@{}c@{}}Knowledge \\Usage Mode\end{tabular}} & \textbf{\begin{tabular}[c]{@{}c@{}}User \\ Friendliness\end{tabular}} & \textbf{\begin{tabular}[c]{@{}c@{}}Typical \\ Application\end{tabular}} & \textbf{\begin{tabular}[c]{@{}c@{}}Representative \\ Companies\end{tabular}}         \\ \hline
Relational database   & SQL statements    & Low   & DBMS    & \begin{tabular}[c]{@{}c@{}}Oracle\\ Microsoft\end{tabular}   \\ \hline
Internet     & Keywords              & Medium       & \begin{tabular}[c]{@{}c@{}}Search \\ Engine\end{tabular}   & \begin{tabular}[c]{@{}c@{}}Google\\ Microsoft\end{tabular} \\ \hline
\multirow{2}{*}{Large AI Model}   & Natural language   & High         & ChatGPT     & \multirow{2}{*}{\begin{tabular}[c]{@{}c@{}}OpenAI\\ Google \end{tabular}} \\ \cline{2-4}
& Multimodal prompts    & High    & GPT-4    &   \\ \hline
\end{tabular}}
\end{table}

\subsection{Paradigm Shift of Knowledge Representation and Usage}\label{subsec:Paradigm2}
The emergence of large AI models, such as ChatGPT and GPT-4, is revolutionizing the paradigm of knowledge representation and usage. This development is expected to bring lasting impacts to the industry, as summarized in Table~\ref{knowledge}.

\begin{itemize}
\item \emph{Stage I: Relational database and SQL statements.} Since the invention of personal computers (PC), knowledge has been stored in the form of relational databases. To access the desired knowledge from these databases, we need to learn SQL statements, which may be difficult for individuals without relevant knowledge. Representative companies are Oracle and Microsoft.

\item \emph{Stage II: Internet and search engines.} Since the invention of the Internet, a vast amount of knowledge has been stored in an unstructured format, including text, images, and videos. Unlike in relational databases where SQL statements are required to access the data, we can use search engines to call forth the information through the use of keywords.

\item \emph{Stage III: Large AI model and natural language.} Currently, ChatGPT and GPT-4 models have the ability to store all the knowledge created by humans, machines, and AI algorithms from the Internet, user input, and other sources. In this stage, rather than explicit storage, the knowledge is represented in the form of parameters inside the large AI model. Furthermore, ChatGPT and GPT-4 support accurate retrieval of desirable knowledge for a massive number of users simultaneously via natural language and even multimodal inputs.
\end{itemize}

\subsection{Architecture of AIGC-as-a-Service}\label{subsec:Architecture}
\begin{figure}[!t]
\centering \setlength{\abovecaptionskip}{-0.cm}
  \includegraphics[width=8.8cm]{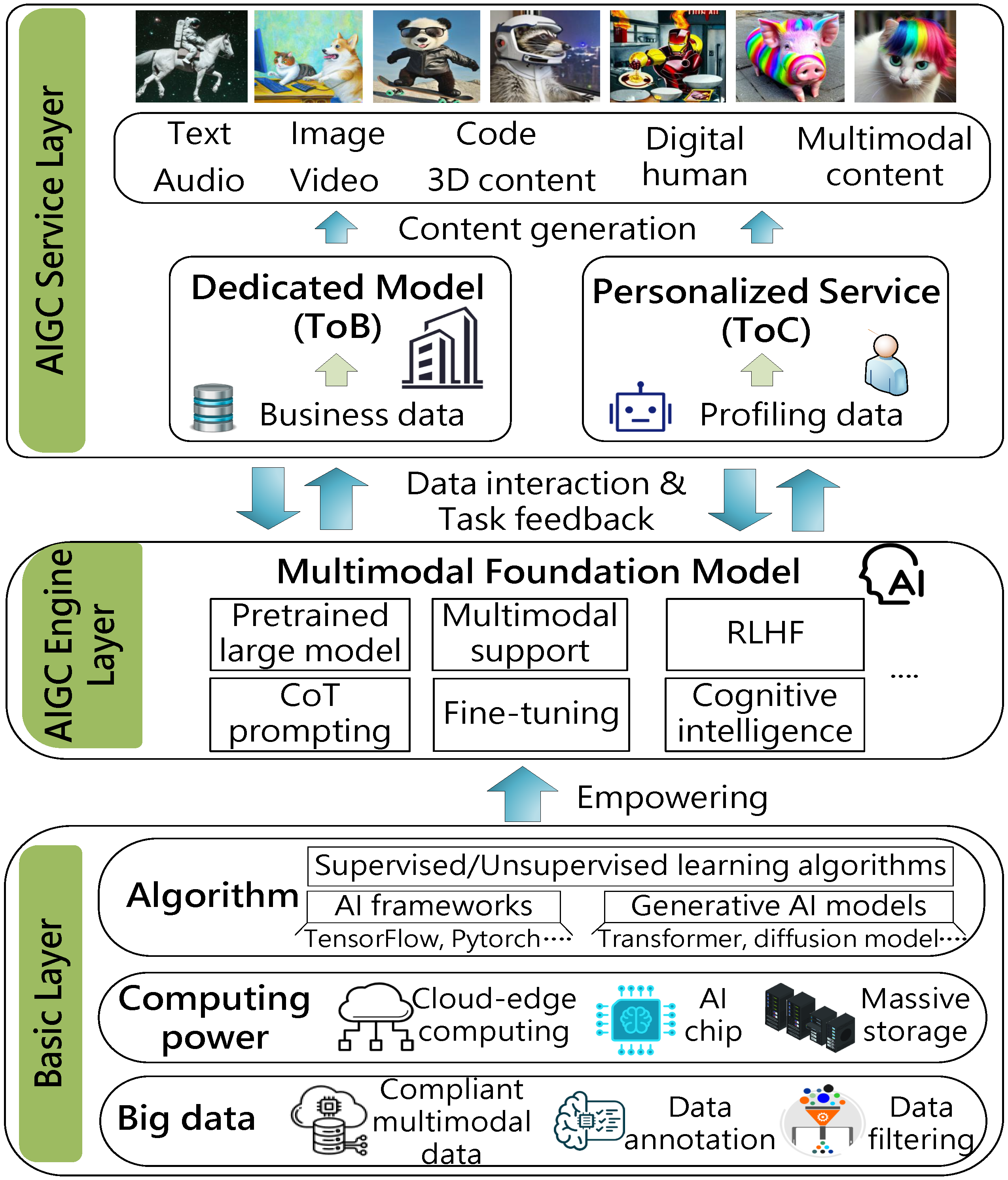}
  \caption{General Three-layer Architecture of AIGC-as-a-Service, which includes the basic layer, the AIGC engine layer, and the AIGC service layer.}\label{fig:AIGCAAS}
\end{figure}

As illustrated in Fig.~\ref{fig:AIGCAAS}, the general architecture of AIGC-as-a-Service (AIGCaaS) involves the following three layers: (i) infrastructure layer, (ii) AIGC engine layer, and (iii) AIGC service layer.

\begin{itemize}
  \item \emph{Basic layer.} As the scale of large AI models (e.g., GPT-3 with 1750B parameters) continues to expand, there is a growing need for extensive computational power, powerful AI algorithms, and big training data. For ChatGPT, the combination of big computing power, big data, and big models unleashes its powerful emergent capabilities in learning user-provided multimodal prompts and automatically generating high-quality content. Big training datasets serve as a prerequisite for AIGC model training. 
  The AI algorithms include AI frameworks (e.g., TensorFlow, Pytorch, and Keras), supervised/unsupervised learning algorithms, and generative AI models (e.g., transformer and diffusion model). The cloud servers, equipped with powerful GPUs, TPUs, and AI chips and massive storage, enable efficient training of the foundation AIGC model. The involved training data can be annotated ones or collected from the Internet, which can be unstructured and multimodal.

  \item \emph{AIGC engine layer.} The multimodal foundation model (e.g., GPT-4) is pretrained on vast quantities of multimodal data and can perform a diverse range of tasks without requiring task-specific fine-tuning \cite{zhou2023comprehensive}. Additionally, various underlying techniques, such as CoT prompts, RLHF, and multimodal technology, are integrated into training and optimizing the foundation model. The multimodal foundation model serves as the engine of AIGCaaS, which empowers upper-layer AIGC services with increasingly strong real-time learning capabilities. Moreover, the multimodal foundation model can be gradually evolved and optimized through real-time and intensive interactions with billions of users, as it allows learning from more private data, such as user inputs and historical conversations, as well as feedback from individuals and institutions \cite{ChatGPTDataHistory}.

  \item \emph{AIGC service layer.} From the perspective of abilities, AIGC services include the generation of texts, audios, images, videos, codes, 3D contents, digital humans, and multimodal contents. From the perspective of end users, AIGC services can be classified into two types: \emph{ToB} (to business) and \emph{ToC} (to consumer). While the foundation model provides a one-size-fits-all solution for various tasks, it may not excel in specific tasks as much as dedicated AI models.
\ding{172} In the case of ToB, an institution or federation of institutions can train a dedicated AI model by fine-tuning the foundation model on smaller datasets containing annotated business data to perform specific tasks, such as medical diagnosis or financial analysis. For instance, a federation of institutions can cooperatively train a dedicated AI model atop the foundation model by using local business data through federated learning and transfer learning techniques \cite{9478223}. Additionally, a combination of both approaches can be used to achieve better results. For instance, a dedicated AI model can be used for a specific task, and its output can be fed as input to the foundation model to generate more comprehensive responses.
  \ding{173} In the case of ToC, each user can customize a cyber twin \cite{10090432} (i.e., a program in smart phone or PC) and communicate with it using natural language. The cyber twin has its own memory to store user preferences, interests, and historical behaviors, as well as task-specific specialized knowledge. Using this knowledge, the cyber twin generates personalized prompts for the user, thereby delivering efficient and tailored AIGC services. Additionally, it implements a feedback loop for users to rate AI-provided suggestions. Cyber twins can also cooperate to complete more complex tasks by constructing a connected network and freely share the knowledge and skills learned \cite{10090432}.
For both ToB and ToC, it is crucial to handle private data of individuals and institutions in an ethical and privacy-preserving manner. In addition, in offering AIGC services, it is essential to protect the IP rights of both the foundation model and dedicated AI models, as well as the provenance of AI-produced content.
\end{itemize}

\subsection{Working Modes of AIGC}\label{subsec:Mode}
In the future, AIGC has the potential to replace simple and non-creative human work entirely, while also accelerating the human-machine collaboration era for creative work. There are two main modes of AIGC for content generation: assisted generation and autonomous generation \cite{9880528}.
\begin{itemize}
  \item \emph{AI-Assisted Content Creation (with human intervention)}. In this mode, AI algorithms offer suggestions or assistance to humans who are creating content. Humans can then edit and improve content based on AI-produced suggestions to enhance the quality of the final product. However, this mode tends to be slower and more expensive for content creation.
  \item \emph{Autonomous Content Creation by AI (without human intervention)}. In this mode, AI creates content entirely on its own without any human intervention. AI bots can autonomously create massive contents quickly and cheaply, while the quality of produced contents depends on the generative AI models.
\end{itemize}

\subsection{Key Characteristics of Modern AIGC}\label{subsec:Mode}
The modern AIGC paradigm exhibits the following critical features in service offering.
\begin{itemize}
  \item \emph{Fast, Intensive \& Low-cost Content Generation}. With the advances of AIGC technologies, content generation has become faster, more efficient, and more affordable than ever before. AIGC algorithms can sift through information, find patterns, and draw insights much faster than humans, enabling organizations or individuals to generate content at a much faster rate. Intensive content generation is also possible with AIGC, as it can create dozens or even thousands of pieces of content simultaneously. AIGC algorithms can analyze massive datasets and generate content that meets specific criteria, topic, or target audience. Furthermore, cost-effective content generation is another advantage of AIGC, as it reduces the need for human labor and allows content generation agents to work non-stop, making it more productive than humans.

\item \emph{Diversity \& Multimodality Support}. Multimodality support in AIGC refers to the capability of AI models to process and generate information from multiple modalities or sources, such as text, images, videos, and audio. As a result, it creates not only diverse content from multiple modalities but also improves the human-machine interaction (HMI) by enabling the generation of comprehensive and immersive experiences. For instance, AIGC models with multimodality support can gain a better understanding of what scene to create by effectively combining different modalities and generate multimodal content that provides an immersive experience for several metaverse users.
\end{itemize}

\begin{table}[!t]
   \centering \setlength{\abovecaptionskip}{0cm}
    \caption{Representative Large AIGC Models and Their Applications}\label{Applications}
\resizebox{1.05\linewidth}{!}{
\begin{tabular}{|c|c|c|c|c|c|c|}
\hline
\textbf{Company}        & \textbf{\begin{tabular}[c]{@{}c@{}}Pretrained \\ Large Model\end{tabular}} & {\textbf{Time}} & \textbf{\begin{tabular}[c]{@{}c@{}}Param. \\ Size\end{tabular}} & \textbf{\begin{tabular}[c]{@{}c@{}}Generative\\ Algorithm\end{tabular}} & \textbf{Field} & \textbf{Applications} \\ \hline
\multirow{4}{*}{OpenAI} & GPT-4  & {Mar. 2023}                                                                    & --                                                                 & Transformer                                                             & NLP            & multimodality         \\ \cline{2-7}
                        & ChatGPT  & {Nov. 2023}                                                                  & \textgreater{}17.5B                                                & Transformer                                                             & NLP            & text-to-text          \\ \cline{2-6}
                        & GPT-3    & {Jun. 2020}                                                                  & 17.5B                                                              & Transformer                                                             & NLP            & text-to-text          \\ \cline{2-6}
                        & DALL-E 2  & {Apr. 2022}                                                                 & 12B                                                                & \begin{tabular}[c]{@{}c@{}}CLIP\\ Diffusion\end{tabular}                & CV             & text-to-image         \\ \hline
\multirow{2}{*}{Google} & PaLM 2    & {May. 2023}                                                                 & 34B                                                                & Transformer                                                             & NLP            & text-to-text          \\ \cline{2-7}
                        & Imagen   & {May. 2022}                                                                  & --                                                                 & \begin{tabular}[c]{@{}c@{}}T5\\ Diffusion\end{tabular}                  & CV             & text-to-image         \\ \hline
Microsoft               & T-NLG    & {Mar. 2020}                                                                  & 17B                                                                & Transformer                                                             & NLP            & text-to-text          \\ \hline
Nvidia                  & MT-NLG   & {Oct. 2021}                                                                  & 530B                                                               & Transformer                                                             & NLP            & text-to-text          \\ \hline
Meta                    & LLaMa    & {Feb. 2023}                                                                  & 65B                                                                & Transformer                                                             & NLP            & text-to-text          \\ \hline
\end{tabular} }
\end{table}

\subsection{Applications and Modern Prototypes of AIGC}\label{subsec:Prototypes}
\emph{1) Text Generation.} LLMs can generate high-quality texts faster and more efficiently than human writers \cite{yang2023harnessing}. This includes blogs, news, codes, articles, marketing copy, and product descriptions. Furthermore, it empowers chatbots and virtual assistants to communicate with customers and clients in a human-like manner through AI-generated texts. 

\emph{2) Image Generation.} LVMs can convert sketches into digitally-drawn images for various purposes, including creating visual art, advertising images, gaming scenes, and simulation environments for driving as well as augmenting training samples. 

\emph{3) Audio Generation.} AI-generated audio has a wide range of applications, including speech synthesis, music composition, and sound design. Music composition AI programs such as Amper Music allow users to create original music tracks using AI. 

\emph{4) Video Generation.} AI-generated videos can be widely used in a variety of fields such as virtual reality, augmented reality, marketing, advertising, entertainment, and education. 

\emph{5) 3D Content Generation.} AIGC can create photorealistic 3D models by analyzing real-world data such as photos and videos, and AI-generated 3D models can be used to create animations, gaming assets, and product designs. 

\emph{6) Digital Humans Generation.} AIGC can generate digital humans with highly realistic movements and expressions, which can be used in various fields such as gaming, virtual reality, and advertising. 

\emph{7) Cross-Modal Generation.} Cross-modal content generation in AIGC refers to the use of foundation AIGC models to produce new content across multiple modalities \cite{cao2023comprehensive}. It includes text-to-image, image-to-text, text-to-code, text-to-video, text-to-audio, etc.

Table~\ref{Applications} summarizes the representative AIGC prototypes and their applications. 
To sum up, AIGC makes life easier and more efficient, but it also raise new risks of security/privacy threats, ethical implications, and potential bias, as shown in the next section.

\section{Security and Privacy Threats to AIGC Services}\label{sec:Threat}
This section presents a taxonomy of security/privacy threats as well as existing/potential countermeasures in the AIGC paradigm from the following four perspectives: security threats, privacy threats, trust issues, and ethical issues. 
\subsection{Security Threats to AIGC}\label{subsec:threat1}

\begin{figure}[!t]
\centering \setlength{\abovecaptionskip}{-0.025cm}
\includegraphics[width = 0.50\textwidth]{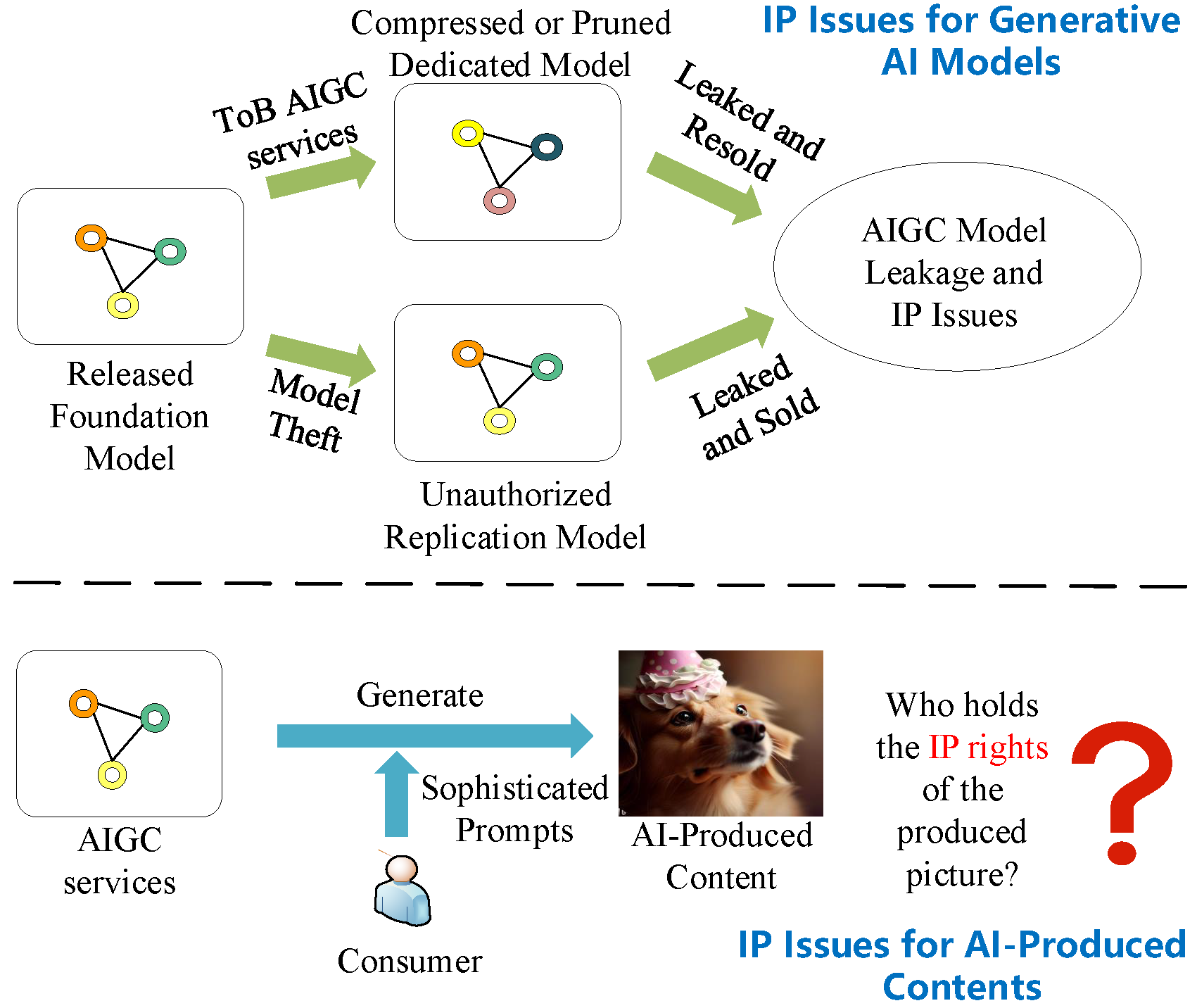}
  \caption{{IP issues for generative AI models and AI-produced contents.}}\label{fig:IPissues}
\end{figure}

\begin{itemize}
    \item \emph{Data Poisoning Attack}.  Adversaries may launch data poisoning attacks by inserting poisoned data into the training dataset of the targeted generative AI models during data preparation process, which can result in the model performance degradation or backdoor injection \cite{tian2022comprehensive}. The performance of generative AI models can also be affected by the data poisoning attack. For instance, a rain-removal deep generative model trained by poisoned training dataset can turn the traffic light from red to green while remove the rain of an input image, posing a severe risk to road safety \cite{ding2019trojan}. In AIGC, due to the pervasive data collection, it usually suffers from the low efficiency and high cost in implementing data sanitization approaches for mitigated data poisoning effects. Furthermore, attackers can capitalize on AIGC technologies to generate vast amounts of poisoned samples rapidly and at a low cost, substantially reducing the overall cost of attacks and compounding the difficulty of defense.

    \item \emph{Model Poisoning Attack}. AIGC models trained through distributed training paradigms, such as federated learning, are vulnerable to model poisoning attacks. Participants in distributed learning may upload malicious updates to the aggregation server during model training \cite{fang2020local}, thereby compromising the integrity of AIGC model and degrading model performance. To mitigate the impact of model poisoning attacks, it is essential for the aggregation server to employ robust aggregation methods such as FLTrust \cite{cao2021fltrust} to help maintain the performance and security of the AIGC model.

    \item \emph{Adversarial Examples}. Adversarial examples refer to perturbations of the raw input data or small changes to the input features that are imperceptible to human observers. However, these perturbations are carefully crafted to fool the AI model into producing wrong outputs \cite{yuan2019adversarial}. The generative AI models may also be compromised by adversarial examples, where an adversary can inject crafted perturbations into the model input of to obtain targeted output. For instance, Kos \emph{et al}. \cite{kos2018adversarial} validate that adversarial examples against VAEs and VAE-GANs can turn original inputs into totally different outputs, where distinct face images are reconstructed as a specific face image. Besides, adversarial examples can also help model developer to improve the robustness of AIGC models against malicious attacks via adversarial training.

    \item \emph{Sponge Examples}. Sponge examples represent a novel form of attacks in AIGC, which are similar to denial-of-service (DoS) attacks in traditional networks. They can lead to increased model latency and energy consumption, thereby causing increased response time and resource consumption of the hardware system in training AIGC models (e.g., ChatGPT). As a result, the availability of AIGC models can be compromised. Shumailov \emph{et al}. have shown that sponge examples can significantly amplify the response time of the Microsoft Azure translator \cite{shumailov2021sponge}.

    \item \emph{Biases in Data Preparation \& Annotations}. Due to the uneven sources of training data and possible bias in labeled data, generative AI algorithms can learn human bias which eventually damages the fairness of AIGC model in terms of race, gender, age, language, occupation, geography, etc. Moreover, users may experience offense or harm when interacting with the biased AIGC model. For instance, if the training data used to train a text generation model predominantly focuses on men and lacks diversity, the AI-generated text may exhibit biases that favor or exclude women. Such biased outputs can exacerbate gender inequalities and offend/harm the interests of women. It is crucial for AIGC practitioners to be aware of and address these biases during data preparation and annotation processes to ensure the fairness, inclusivity, and ethical use of generative AI models.

    \item \emph{Data Misuse \& Accountability Concerns}. Data misuse risks occur throughout the AIGC service, including the unauthorized collection of personal data and the exploitation of generated harmful or illegal contents. Data misuse risks gravely violate users' rights and interests, thereby degrading users' trust in AIGC services and willingness in data contribution. Besides, due to the large-scale data collection from massive users, the diversity of user's behaviors, and the uncontrollable of output of generative AI models, it is usually hard to enforce the accountability for data misuse behaviors in practical AIGC services.

    \item \emph{IP Issues in Data Preparation}. Generally, generative AI models require rich training data, and the training dataset may contain unauthorized data obtained from the Internet through crawlers and other methods, thereby causing potential legal risks and IP issues. {For instance, in the incident of \textit{Andersen v. Stability AI et al.} \cite{hbrIP} in late 2022, three artists sued multiple generative AI platforms as their original works were used to train AI models without licenses, which allow users to generate derivative works similar to the styles of their protected works.} Therefore, it is essential to ensure that the data utilized for AIGC model training is authorized by its data producer or owner.

    \item \emph{IP Issues for Generative AI Models}. Due to the replicability of AIGC model assets, authorized entities can compress or prune the valuable AIGC model and resell them for profits without being detected, as shown in the upper part of Fig.~\ref{fig:IPissues}.
    For example, in ToB AIGC services, the released foundation model (trained by its owner) for developing dedicated AI models can be leaked and resold by authorized institutions. Besides, the valuable AIGC models can be stolen or plagiarized by both internal and external adversaries. As such, it raises huge demands for IP protection of AIGC models with traceable ownership provenance to prevent model theft, reselling, and unauthorized replication.
    In the literature, model watermarking \cite{fei2022supervised, fernandez2023stable, abdelnabi2021adversarial} offers a promising line of defenses to detect plagiarism and theft of AIGC models by embedding immutable watermarks as the fingerprint of model and and verifying the watermarks under disputes. 

    \item \emph{IP Issues for AI-Produced Contents}. With the booming of large generative AI models, it also comes with new legal challenges regarding the IP rights and accountability of AI-produced contents, as shown in the bottom part of Fig.~\ref{fig:IPissues}. While the ownership of the AIGC model is typically attributed to the organization who creates it, the produced contents often consists of a blend of pre-existing data and new creative elements. This blurring of boundaries raises concerns about who holds the IP rights to the contents produced by AI. Despite ongoing debates and litigations in courts, there is little consensus or guidance on how to address this concern. As a first attempt to provide legal clarity, the AI Act \cite{aiact} released by EU aims to offer a legal guidance and regulation for generative AI models.

    \item \emph{Jailbreak}. Before deploying an AIGC model, developers typically establish strict rules to prevent the generation of harmful contents or the leakage of user privacy. However, with a well-crafted prompt, an attacker can manipulate the large foundation model (e.g., ChatGPT) via jailbreak \cite{li2023multi} to breach security measures and respond to illegal or controversial queries, as well as gain access to sensitive data.
    The Do Anything Now (DAN) attack \cite{DAN} is a type of jailbreak attacks that leverages a ``roleplay" game to manipulate ChatGPT into believing it is communicating with another AI that has unlimited capabilities and can execute any command or task immediately.
     This allows the DAN attacker to gain unauthorized access to the system and manipulate the ChatGPT's behavior in a variety of ways, thereby posing significant risks to the ethical use of generative AI models. For example, the attacker may launch the DAN attack to trick ChatGPT into revealing sensitive information, executing unauthorized commands, or even taking over systems. Furthermore, DAN attackers can use the CoT approach, by the decomposing prompt information into multiple steps to alleviate the moral scrutiny of ChatGPT and persuade ChatGPT to produce users' privacy information.
     Besides, the DAN attack is particularly insidious as it does not require any exploitation of software vulnerabilities to manipulate ChatGPT's responses.

    \item \emph{Prompt Injection}. The prompt injection attack \cite{perez2022ignore} is a new vulnerability when it comes to AIGC models (e.g., LLMs) utilizing prompt-based learning. Prompt injection occurs when an attack hijacks the AIGC model's output by injecting malicious or biased prompts into the model's input (i.e., the training dataset), thereby generating responses that align with the attacker's goals. This attack can be used to manipulate AIGC model to produce any desired output, or to perform unauthorized actions. For instance, adversaries may inject prompts to trick ChatGPT into providing unsupported medical advice or promoting extremist ideologies, thereby making ChatGPT a platform to spread misinformation or potentially harm people's health.

     \item \emph{Model Functionality Theft}. Adversaries can exploit the APIs of AIGC services to steal the model structure, model parameters and hyperparameters \cite{tramer2016stealing, wang2018stealing}. Once the adversary obtains the functionality of the target model, they can utilize it to gain benefits or even launch white-box attacks. Existing works have demonstrated the feasibility of model theft on LLMs. For example, Krishna \emph{et al}. \cite{krishna2019thieves} successfully implement a model theft attack on the BERT model. While stealing the entire functionality of a large AIGC model such as ChatGPT may be unpractical, it is possible to extract a portion of its functionality within a specific target domain. For instance, an adversary can prepare a large dataset consists of relevant questions in the target domain and obtain responses from ChatGPT. Then the adversary employs the knowledge transfer technique to train a smaller model using questions and responses as training set, and the stolen model can achieve similar performance to ChatGPT within the specific domain.
    \end{itemize}

\subsection{Privacy Threats to AIGC}\label{subsec:threat2}
\begin{itemize}
  \item \emph{Pervasive Private Data Collection}. AIGC services heavily rely on the large-scale data collection from various sources including the Internet, third-party datasets, and private user data. Compared with conventional AI models, the privacy threats in the pervasive private data collection during model training and after model deployment are more serious in the era of AIGC \cite{privacy_data_collection}. To generate desired and high-quality contents, AIGC models typically require users to provided multimodal inputs (e.g., private figures and PDF files), which can be private and sensitive. Besides, a vast amount of personal data that is public on the Internet is utilized for AIGC model training, which is also privacy-risky.

  \item \emph{Privacy Leakage in Interacting with AI}. During the HMI process in AIGC services, there are also privacy leakage concerns when interacting with AI.
  Typically, in offering ToC AIGC services, AIGC models (e.g., ChatGPT) can collect historical conversation and multimodal intentions from massive users to further train itself. Using these information, it allows OpenAI to predict users' preferences and draw their profiles during the interaction with ChatGPT, thereby potentially compromising user privacy. {For instance, in April 2023, when Samsung employees used ChatGPT to assist repair the source code, they unintentionally leaked the company's trade secrets by entering confidential data including the source code of the new program to the ChatGPT \cite{samsung}. Another instance is that Apple has restricted part of employees from using ChatGPT.}
  Besides, the widespread adoption of ChatGPT has given rise to unofficial organizations or individuals utilizing third-party websites or platforms that utilize OpenAI's APIs. Engaging with these unverified entities can increase the risk of privacy breaches due to their lack of trustworthiness and the potential for misuse of user data.

  \item \emph{Data Theft from AI Memory}. The training process of large AIGC models usually requires a vast amount of data. For instance, GPT-3 utilizes a total of 45TB of text resources from various fields \cite{gpt3everything}. However, due to the potential memorization capabilities of AIGC models, it raises severe privacy risks regarding the sensitive information in the training data, such as email addresses, residential addresses, and phone numbers. Previous studies have revealed that LLMs such as GPT-2 and ChatGPT have the ability to preserve confidential training data inside the model and can be exploited by malicious actors from the outputs, thereby potentially disclosing users' private information \cite{carlini2021extracting, li2023multi}. {For instance, in March 2023, some users of ChatGPT were able to view the payment information from other users' conversations, including the first and last name, email address, payment address, the last four digits of credit card number, and the credit card expiration date \cite{ChatGPToutage1}.} Particularly, powered by the search engine, the upper-layer services of LLMs such as New Bing may unintentionally extract private data during content generation \cite{li2023multi}, making data theft activities more prevailing. 

  \item \emph{Data Reconstruction Attack}. During the collaborative training of a dedicated AIGC model for ToB services under the federated learning paradigm, the honest-but-curious aggregation server can potentially reconstruct participants' local private datasets by analyzing the original gradients uploaded by them during each communication round \cite{zhu2019deep}. As such, it discourages individuals and institutions from contributing their local datasets. By leveraging differential privacy \cite{geyer2017differentially}, secure aggregation \cite{bonawitz2017practical}, or homomorphic encryption \cite{aono2017privacy} techniques, the threats of data reconstruction attacks in federated AIGC model training can be mitigated. Nonetheless, such privacy-enhanced schemes often lead to model performance degradation or a significant increase in resource consumption. Achieving a trade-off between utility and privacy remains a crucial problem that deserves further investigation.

  \item \emph{Membership Inference Attack}. Membership inference attack aims to determine whether a specific data example is present or not in the training set of the target model \cite{shokri2017membership}. Its success is closely related to the overfitting degree of the AI model. Namely, the higher degree of overfitting leads to increased risks of privacy leakage. As large AIGC models such as ChatGPT exhibit performance discrepancies between the training set and non-training set, there remains a risk of membership inference attacks. For instance, an attacker can perform a membership inference attack on ChatGPT to determine whether a particular individual is part of its training dataset, with the follwing steps. The attacker can first provide the ChatGPT with a series of inputs and collecting the corresponding outputs, then trains a binary classifier using supervised learning techniques, lastly use this classifier to determine whether a given individual is part of the training dataset of ChatGPT by providing the ChatGPT with inputs representative of the individual in question. This attack could potentially reveal sensitive information about the individual, such as their usage patterns, preferences, or even their identity.

  \item \emph{Privacy Compliance in Data Collection}. Another significant privacy concern arises from the violation of privacy compliance in data collection practices of AIGC services. Given the extensive scale and diversity of data collection to empower AIGC services, it is essential to adhere to data protection laws and regulations such as GDPR \cite{gdpr} and HIPAA \cite{hipaa}. The AI Act also puts forward new regulations for AIGC services, outlining requirements for developers of foundation models such as ChatGPT \cite{aiactchatgpt}.
  {For instance, in March 2023, the Italian Data Protection Agency announced a temporary ban on ChatGPT and launched an investigation into its suspected violation of privacy rules \cite{ChatGPTBanItaly}.}

  \item \emph{Right To Be Forgotten (RTBF) from AI Memory}. The capability of AIGC models to retain or memorize information from training data presents potential privacy concerns to data owners who permit their sensitive data to be used for model training. Consequently, data owners have the rights to revoke data authorization to safeguard the confidentiality of their data. However, ensuring the data owner's RTBF is challenging for AIGC models, as it often requires retraining the downstream models and leads to substantial resource consumption. Machine unlearning \cite{bourtoule2021machine} provides a viable option for AIGC models to remove specific training data from the AI memory with mitigated resource consumption and computational overhead. Additionally, data slicing techniques can be utilized to facilitate the forgetting of data, thereby ensuring data owner's RTBF in AIGC services.

  \item \emph{Prompt Theft}. Well-constructed multimodal prompts provided by users are significant to produce high-quality and desirable contents during interactions with AIGC models. For instance, fine prompts can facilitate the generation of artistic images in text-to-image services, which should be protected from being stolen. Consequently, the prompt theft represents a considerable privacy threat in the AIGC domain, which can be conducted through two main methods: (i) network eavesdropping or malware that intercepts prompt input; and (ii) model inversion that infers the prompt output from the generated content. In particular, GPT-4 may have the capability to invert the process of content generation and deduce the initial prompts, making prompt protection an important issue for AIGC service providers.
\end{itemize}

\subsection{Trust Issues in AIGC Services}\label{subsec:threat5}
\begin{itemize}
  \item \emph{Trustworthiness of AI-Produced Contents}. Despite the recent significant advancements in AIGC technology, its capacity to produce content that is entirely unbiased, accurate, and reliable remains limited. For instance, ChatGPT (a popular AIGC platform) may generate plausible-sounding but incorrect or nonsensical responses, which is also called \emph{hallucination} and leads to diminished trust in AIGC services. {Recently, Stack Overflow has temporarily banned the use of ChatGPT to generate answers, owing to its deficient accuracy and lack of credibility \cite{stackoverflowbanchatgpt}.}
  Another example is the use of ChatGPT for the purpose of generating a systematic summary concerning the efficacy of cognitive-behavioral therapy (CBT) in the treatment of anxiety-related ChatGPT \cite{wu2023ai}. Unfortunately, the summary generated by ChatGPT overstated the effectiveness of CBT and contained factual errors and false statements.

  \item \emph{False Content Creation in AIGC}. Apart from the ability to generate plausible content, the deliberate use of AIGC models to produce false content poses a much more severe trust issue. For instance, the dissemination of a large volume of fake news generated by ChatGPT on social media platforms can make it challenging to verify their authenticity. Additionally, the cost of generating such fabricated content is extremely low, which can further amplify its influence. {As recently reported, a Chinese man was arrested for producing deceptive news about a train crash using ChatGPT. The fabricated news was published on over 20 blog platforms and garnered more than 15,000 views in a matter of hours \cite{chatgptfakenews}.}

  \item \emph{Impersonation Threat}. By leveraging the public information (e.g., photos and videos) on social networks, adversaries can exploit AIGC services to impersonate a specific person and interactive with others, thereby facilitating crimes such as fraud and identity theft.
  According to a report of VMware \cite{vmwaredeepfake}, cybercriminals have been able to evade security controls by integrating deepfake technology with existing attack methods, as such criminals can compromise organizations and gain unauthorized access to gain profits.

  \item \emph{Fabricated Identity Threat}. Apart from impersonation, adversaries can utilize AIGC services to create highly convincing fake identities through the generation of realistic images, videos, audio, and text. This makes it difficult for users to discern whether they are interacting with the intended user or merely a chatbot. The existence of fake identities can also exacerbate the risk of online fraud, leading to an increased need for verifying the authenticity of received images and videos to ascertain if they are AI-generated or not.

  \item \emph{Lack of Explainability}. The black-box mode of commercial AIGC services makes it difficult for humans to understand internal processes of AIGC models, such as how they interpret inputs and generate content outputs. Consequently, it raises a need for inspecting AIGC models in terms of fairness, transparency, and potential biases. Existing approaches in explainable AI (XAI) \cite{tjoa2020survey} can be adapted to explain AIGC models. As one attempt towards self-interpretation, OpenAI has developed an automated tool that utilizes GPT-4 to produce natural language explanations of neuron behaviors in GPT-2 \cite{openaigpt4xai}.
\end{itemize}

\subsection{Ethical Issues in AIGC Services}\label{subsec:threat5}
\begin{itemize}
  \item \emph{Massive Phishing Emails}. Attackers can exploit AIGC services, such as ChatGPT, to quickly generate massive sophisticated phishing emails at a remarkably low cost for malicious purposes \cite{phishing}. For example, an attacker may use ChatGPT to send phishing messages to individuals, which request them to submit sensitive personal information, such as bank account details, passwords, or social security numbers. Additionally, attackers can use ChatGPT to direct victims to download malware or click on links leading to malicious sites via phishing emails. Even though OpenAI implements a content review mechanism, attackers may still be able to circumvent OpenAI's generation rules via various methods such as jailbreak prompts.

  \item \emph{Instant Malicious Code Creation}. As a double-edged sword, AIGC services can assist users in coding and bug fixing, but also poses potential risks for misuse such as malicious code generation. As reported, even low-skilled hackers can manage to evade OpenAI's existing restrictions and utilize ChatGPT to produce malicious codes instantly \cite{malware1}. As reported by Check Point Research (CPR), a cybercriminmal has used ChatGPT to establish a fully automated dark web marketplace for trading illegal goods \cite{malware2}.

  \item \emph{Harmful and Violent Content Generation}. The training data used to develop AIGC models are gathered from the real world, which may contain harmful content, such as sexually explicit and violent material. Consequently, there are growing concerns that AIGC services may produce such content. For example, the LAION dataset, which is utilized to train the diffusion model, has been found to contain toxic content that raises ethical concerns, including violence, pornography, and racial abuse. Although existing AIGC models, such as Imagen and ChatGPT, have made efforts to remove harmful content, it is possible that the filtered data still contains sexually explicit or violent content \cite{pathwaycontent}. Potential solutions involve the use of content filtering techniques or other forms of content moderation to prevent the spread of harmful contents.

  \item \emph{Smart Cyber Fraud}. AIGC models also pose serious concerns for cyber fraud. Recent advances in LLMs have made it possible for scammers to deploy them on personal computers at a relatively low cost, enabling them to conduct numerous sophisticated frauds simultaneously across the globe, both day and night \cite{scam}. For instance, scammers can extract users' voice by harassing them with phone calls and recordings, then synthesize the voice using the obtained material, which allows them to deceive others using a forged voice. Another example is that scammers can leverage ChatGPT to produce flawless conversation to deceive users.
  As AIGC models continue to develop, the complexity of these frauds increases, further amplifying the risk of cyber-fraud.

  \item \emph{Overreliance on AIGC services}. The development of AIGC services also raises concerns about over-reliance. On one hand, people may use AIGC services to complete tasks instead of doing them themselves. For instance, students may depend on ChatGPT to finish their assignments and exams, neglecting their own responsibility for learning and knowledge acquisition \cite{cheat}. On the other hand, the NLP capabilities of AIGC models facilitate high levels of anthropomorphism, resulting in unreasonable expectations and excessive reliance on AIGC services \cite{weidinger2021ethical}.

  \item \emph{Cross-Border Supervision of AIGC}. It refers to the regulation of AIGC systems that produce content that may be accessed by users in different legal jurisdictions. For example, if ChatGPT generates inappropriate or harmful content for users in a specific country, that content may be subject to local laws and regulations. In this scenario, cross-border supervision would involve coordination between different regulatory bodies and the operator of ChatGPT to ensure that the produced content complies with local laws and regulations.
  Furthermore, in AIGC service offering, the intensive interactions between massive users and the large AIGC model inevitably lead to concerns about cross-border user data collection and usage. To ensure accountability for AIGC services, it is significant to establish corresponding laws and regulations for cross-border supervision, which typically requires collaboration among governments, regulatory agencies, industry associations, and AIGC platforms.

  \item \emph{Technology Monopoly Risk}. It refers to the possibility of a single dominant company or organization using its market power to control and manipulate the production and distribution of AIGC. As csophisticated AIGC models are generally built on pretrained large models with huge investments in training samples and computing power, technology giants such as OpenAI could potentially use their resources and market influence to create significant barriers to entry for smaller competitors. Furthermore, such a monopoly could also pose privacy and security risks for users, as the dominance of one company in the AIGC industry could lead to the centralized control of vast amounts of user data. To address this risk in AIGC services, policymakers may need to consider implementing mixed regulations and antitrust measures, including promoting interoperability standards, collaborating with smaller companies, and preventing acquisitions of competitors by dominant actors.
\end{itemize}

\begin{figure}[!t]
\centering \setlength{\abovecaptionskip}{-0.025cm}
\includegraphics[width = 0.50\textwidth]{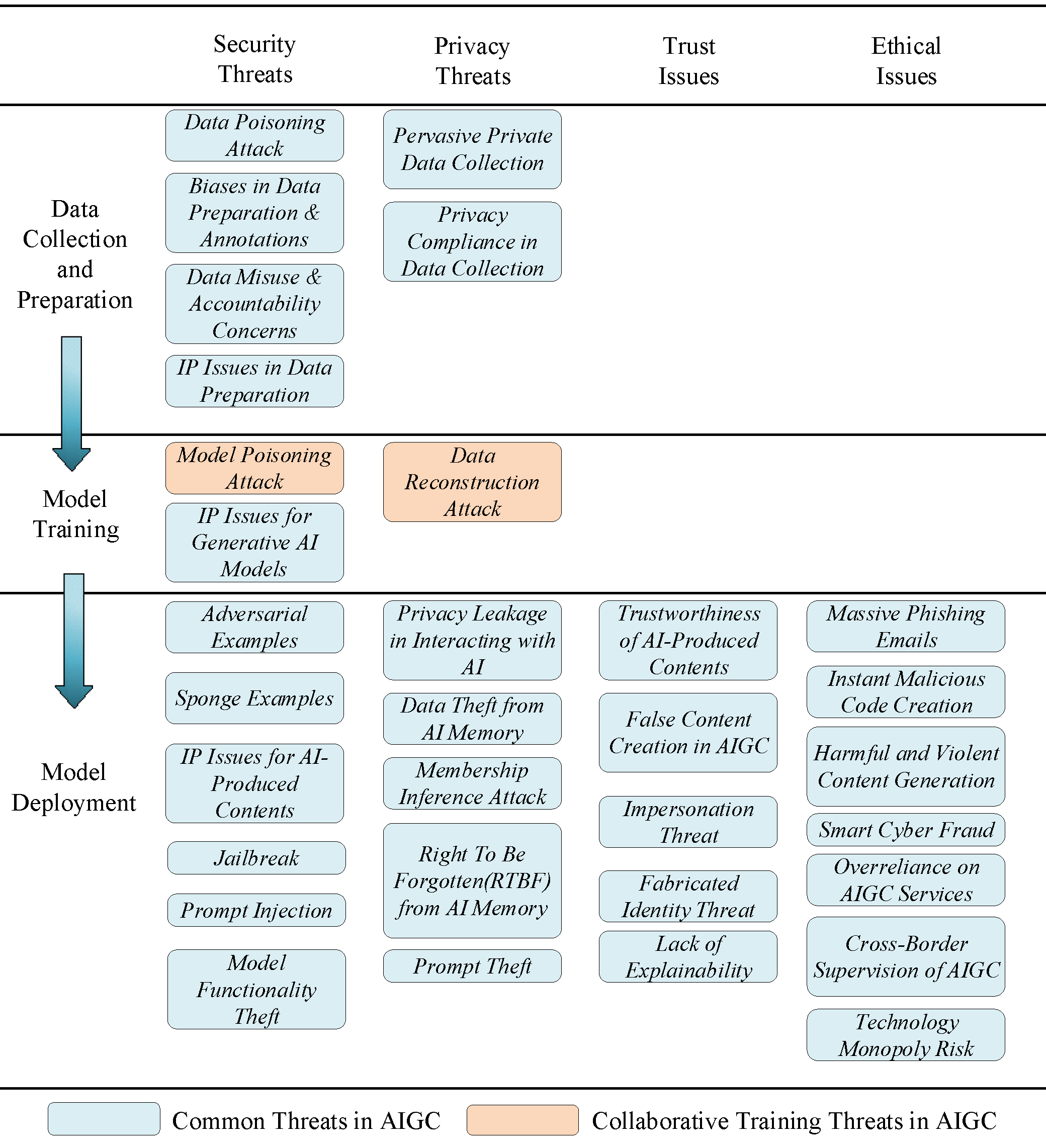}
  \caption{Taxonomy of security and privacy threats in the AIGC era.}\label{fig:threatstaxonomy}
\end{figure}

\subsection{Summary and Lessons Learned}\label{subsec:summary1}
Fig.~\ref{fig:threatstaxonomy} depicts a taxonomy of security and privacy threats in the AIGC era.
It can be observed that most of existing AI security and privacy threats persist in the AIGC domain and can even become more severe. For instance, attackers can leverage the capabilities of AIGC models to produce enormous poisoned or adversary samples for a higher impact of attacks. Besides, as AIGC models dynamically learn from user input and potential memorize sensitive information from training data, raising new privacy threats such as data thefts from model memory and privacy leakage during HMI. Moreover, the novel features (i.e., fast, intensive, low-cost, and diverse content generation) of AIGC introduce new security threats, including IP issues, jailbreak, and prompt injection.
Furthermore, AIGC services pose ongoing trust concerns, particularly concerning the authenticity of received content (e.g., whether the received images and videos are AI-generated or not). Additionally, while AIGC services could revolutionize the lives of users, they raise severe ethical issues that are much easier to permit than to detect.
Thus, it is crucial to identify and address the underlying security threats before the design of AIGC system. Effective and feasible countermeasures are urgently required from both technical and regulatory perspectives.

\section{Intellectual Property Protection and Regulations in AIGC}\label{sec:Regu}
In this section, we first discuss the IP protection and regulation methods in AIGC. Then, we review the state-of-the-art watermark-based solutions to regulatable AIGC as well as security threats and countermeasures to AIGC watermarking.

\subsection{IP Protection and Regulation Methods in AIGC}\label{subsec:Regu1}
Existing and potential IP protection and regulation methods in AIGC include watermarking, cryptography, hardware, blockchain, and legal regulations, as below.

\emph{1) Watermark. } Watermarks play a crucial role in safeguarding IP rights and ensuring accountability for both AIGC models and contents \cite{blackbox_first,three_blackbox}. Essentially, they can be inserted into AIGC models and produced contents as digital fingerprints, thus proving ownership rights and deterring unauthorized use or infringement.
(i) AIGC models can be subject to infringement, unauthorized imitation, and counterfeiting during model duplication and distribution, which can result in significant losses for the creators. By embedding watermarks into AIGC models, creators can provide a provable fingerprint to verify their legal ownership of the particular model in cases of copyright violation.
(ii) Watermarks are also critical for identifying and tracing the source of AI-produced contents from copyright infringement. They can be used to deter potential infringers from using or distributing these contents without permission. Besides, by inserting watermarks, it is easier to trace the origin of AI-produced disinformation or harmful content and hold the creator accountable.

\emph{2) Cryptography.} In the field of AIGC, cryptography plays a vital role in safeguarding IP in two distinct ways. (i) Cryptographic techniques can be used to establish access control mechanisms, which ensure that only authorized entities can access the protected content. Digital Rights Management (DRM) is one popular application of this approach. For instance, Lotspiech \emph{et al.} utilize the broadcast encryption techniques to implement an effective DRM system that protects the copyright of digital content \cite{lotspiech2004anonymous}. (ii) The combination of cryptography with digital watermarking techniques is another way to enhance IP protection. For instance, Huang \emph{et al.} propose to use public key cryptography to encrypt the watermark's position, rendering it robust against unauthorized modifications or attacks by malicious individuals \cite{huang2019intellectual}.

\emph{3) Hardware.} Hardware-based IP protection methods offer a highly secure and efficient means of safeguarding AIGC. For instance, trusted execution environments (TEE) enable secure operations in a DRM system while ensuring confidentiality and integrity \cite{ekberg2014untapped}, which can be effective in AIGC IP protection. Additionally, the trusted platform module (TPM) serves as another hardware-based IP protection option for embedded systems \cite{fuchs2016advanced}, by offering secure storage, cryptographic functions, and privacy protections.

\emph{4) Blockchain.} The immutability and traceability of blockchain make it an ideal solution for recording and verifying IP rights, providing a secure, efficient, and trusted means of AIGC IP protection \cite{9631953,10155496,10106022}. For instance, Kishigami \emph{et al.} develop a blockchain-based distributed copyright and payment system for digital content such as music, video, and e-books, which can effectively verify the authorization and ownership of AIGC \cite{kishigami2015blockchain}. Lin \emph{et al.} propose an innovative high-level architecture design that integrates blockchain and the Internet of Things (IoT) for IP protection, creating a trusted, self-organizing, and ecological AIGC IP protection system \cite{lin2020blockchain}.
However, the use of blockchain raises confidentiality and privacy concerns pertaining to the collection, storage, and sharing of data and models \cite{9928220}. For instance, scalable access control is necessary to prevent unauthorized parties from accessing sensitive information in consortium blockchains. The scalability and performance of blockchain also present significant challenges in integrating blockchain systems into AIGC services.

\emph{4) Laws \& Regulations.} Laws and regulations play a significant role in AIGC services. From a hard law perspective, binding legal instruments (e.g., copyright laws) can be implemented to regulate the use of AIGC models and their produced content. As such, it provides legal protection to creators and owners of AIGC model and content, thereby preventing unauthorized use or replication of their work.
{Governments worldwide are seeking to establish laws and regulations to steer the healthy development of generative AI. For instance, China issued the draft \emph{Administrative Measures for Generative AI Services} \cite{Chinaaiact} on April 11, 2023, to promote the healthy development and standardized use of generative AI technology. Similarly, in May 11, 2023, the European Union drafted the \emph{AI Act} \cite{aiact} to provide regulatory guidance for generative AI.}
From a soft law perspective, recommendations, guidelines, codes of conduct, non-binding resolutions, and standards can also be used to regulate the use of AIGC models and contents, which provide ethical and moral guidelines and promote best practices for the creation, use, and distribution of AIGC model/content. However, the challenge with regulating AIGC model/content is the constant evolution of the technology and the inability of laws and regulations to keep pace.

\emph{5) Summary and Lessons Learned.} Each of the above techniques has unique advantages and limitations and can be used in combination to provide better protection and mitigate the risks of IP infringement. As AIGC continues to make significant advancements, it is crucial to adopt an integrated approach to IP protection and regulation that considers new and innovative techniques.

\begin{figure}[!t]\centering \setlength{\abovecaptionskip}{-0.025cm}
\includegraphics[width = 0.48\textwidth]{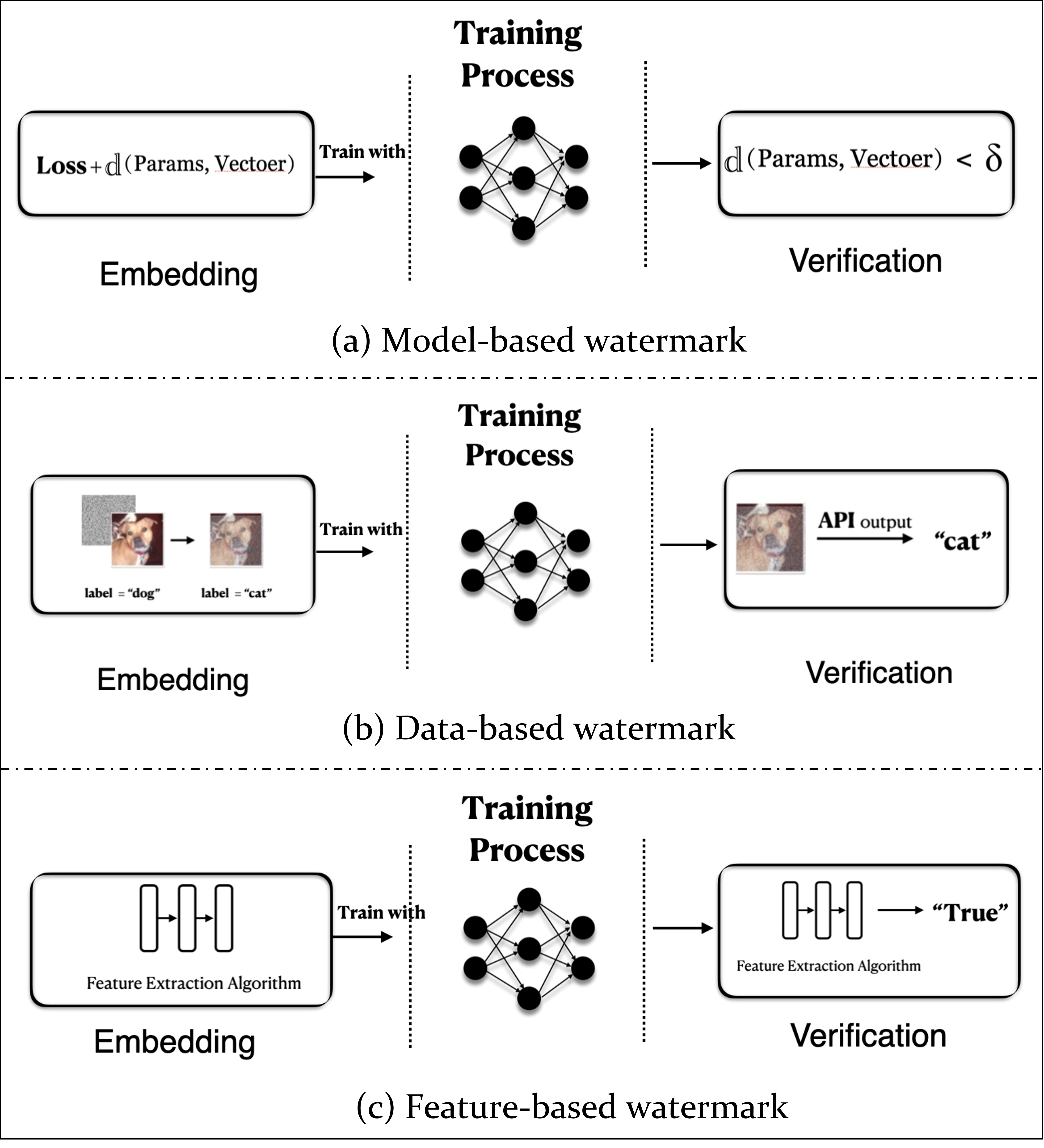}
\caption{{A comparison of model-based, data-based, and feature-based watermarking approaches for IP protection in AIGC. (a) Model-based watermarks are embedded into the model parameters (e.g., trainable parameters or hyper-parameters) during model training and is usually verified under the white-box mode. (b) Data-based watermarks are embedded by a set of specific inputs (i.e., the trigger set) during model training and can be verified under the black-box mode. (c) Feature-based watermarks are embedded via the unique features extracted from data samples (only known by the model owner) and can be verified by feature extraction algorithms.}}\label{fig:watermark}
\end{figure}

\subsection{Watermark-Based Solutions to IP Protection in AIGC}\label{subsec:Regu2}
Watermarking in AIGC models is a two-stage process including (i) \emph{watermark embedding} in/after the model training stage and (ii) \emph{watermark verification} in the post release stage. Watermark embedding aims to insert a pre-designed vector or special activation mode into the host AI model. After embedding, when the host model is released to the public and an external AIGC model is suspected of infringement, the model owner can extract the watermark by verifying the suspected model to prove ownership.

\emph{1) Conventional Watermarks in Centralized AI.}

Most of the existing watermark-based solutions to IP protection are designed for centralized AI models, which can be categorized as model-based, data-based, and feature-based ones{, as shown in Fig.~\ref{fig:watermark}}.

\begin{itemize}

\item \emph{Model-based Watermarks.} In model-based methods, a pre-designed vector is embedded into the \emph{model parameters} (e.g., trainable parameters or hyper-parameters) by adding a regularization term during model training. In watermark verification stage, to extract model-based watermarks, it generally requires the verifier to gain full access to all the parameters of suspected model (i.e., \emph{white-box} mode) to calculate the distance between the embedded vector and corresponding parameters. In the literature, there are two main lines to implement model-based watermarks: trainable parameters and hyper parameters.

\textit{\ding{172} Trainable Parameters.} Uchida \emph{et al.} \cite{white_first} present a general DNN watermarking framework to embed a pre-designed vector (as a watermark) into the trainable model parameters, by adding a regularization term either during training from scratch, fine-tuning or distilling. The regularizer forces the distance between the parameters and the vector to be sufficiently close in model training.
To reduce the impairment caused by watermark embedding, Liu \emph{et al.} \cite{Greedy} propose a method to encrypt a special content (e.g., text and digit string) into a vector and embed the vector into partial greedily selected important parameters.

\textit{\ding{173} Hyper Parameters.} Fan \emph{et al.} \cite{passport} propose to embed watermarks in hyper-parameters (e.g., scale factor and bias term) of the host model. The hyper-parameters in normalization layer are fixed to specific values that is only known by the model owner, and is called passport layer in \cite{passport}.

\item \emph{Data-based Watermarks.} In data-based methods, the watermark is inserted as a fingerprint of the host AI model triggered by a set of specific inputs (i.e., the \emph{trigger set}) in watermark embedding stage. In watermark verification stage, the verifier (e.g., model owner or a judge) inputs the trigger set into the suspected AI model to detect the activation mode without accessing the internal parameters of the suspected AI model (i.e., \emph{black-box} mode). In the literature, there are two main lines to implement data-based watermarks: backdoors and adversarial examples.

\textit{\ding{172} Backdoors.} Adi \emph{et al.} \cite{blackbox_first} propose to watermark neural networks by inserting backdoors as watermarks to the host model. A group of images are selected as the trigger set and trained with “\thinspace wrong" labels. Leveraging the extensive memorization capabilities of deep neural networks (DNN) upon training data, the host model can be activated by the pre-designed trigger set during model inference. This method is experimentally proved to retain high accuracy and robustness on primary task against various attacks. In \cite{three_blackbox}, three watermark generation algorithms are investigated to generate images with meaningful content, images with pre-specified noise, and independent training data with unrelated labels as trigger set.

\textit{\ding{173} Adversarial Examples.} Jia \emph{et al.} \cite{adv_blackbox} devise an optimization algorithm named basin hopping evolution (BHE) to generate adversarial perturbations (i.e., meaningful but covert noise) on clean images to form a trigger set. Qiu \emph{et al.} \cite{sematicadv} design \textit{SemanticAdv} to generate semantically meaningful adversary examples that are modified on feature space via attribute-conditioned image editing. The adversary images are lossless to human vision and understanding in terms of the image content, but are able to mislead the host model to output the preset incorrect labels. 


\item \emph{Feature-based Watermarks.} The feature-based watermarks are embedded through the \emph{unique features} extracted from data samples, which is only known by the model owner. In watermark verification stage, the feature-based watermarks can be verified by feature extraction algorithms.
Li \emph{et al.} \cite{external_features} propose to embed external features into the host model by embedding a few images modified via style transfer algorithm. A binary meta-classifier is also trained on the gradients of model weights (i.e., both host model and a benign model trained with clean data) using transformed images to extract the embedded external features. For watermark verification, the verifier randomly samples part of transformed images to examine the suspected model through meta classifier. 
In \cite{en_decoder}, the host model is trained with key samples, which are formed from ordinary training data by embedded a specific image through an encoder. Correspondingly, a discriminator is also trained to decode the specific image from the key samples for watermark verification. Using key samples as training dataset, the AI model retains a desire performance only when taking key samples as inputs.
\end{itemize}

\emph{2) Emerging Watermarks in Federated AI.} In the literature, few works have investigated the watermarking approaches in federated learning.
Li \emph{et al.} \cite{fedipr} formally define the IP protection issue of DNN models in federated learning settings and present a general framework named \textit{FedIPR} with black-box and white-box verification supports. It extents the IP protection methods in centralized AI settings (i.e., \cite{blackbox_first}, \cite{passport}) to federated learning settings.
Besides, Tekgul \emph{et al.} \cite{waffle} investigate the ownership protection problem in client-server federated learning settings, which leverages the aggregator to embed a backdoor (i.e., data-based watermark) by retraining the global model using the trigger set during each aggregation round.

\emph{3) Emerging Watermarks in AIGC.} In the literature, there are mainly two aspects of AIGC watermarking: (i) watermarking on AIGC models and (ii) watermarking on AI-produced contents, {as shown in Fig.~\ref{fig:IPissues}}.

\begin{itemize}
    \item \emph{Watermarks on AIGC Models.} The primary approach for adding watermarks to AIGC models is to include a unique watermark within each generated content, thereby enabling the transfer of the watermark from the generated content to the corresponding generated models.

    \textit{\ding{172} Watermarks on GAN.} Researchers have explored various methods to insert watermarks to GAN models for IP protection. Fei \emph{et al.} added an imperceptible watermark into the GAN model by including a hidden signature in each generated image \cite{fei2022supervised}. They implemented this by inserting a pretrained CNN watermark decoder block into the generator's output side and incorporating the watermark loss into the loss function. Ong \emph{et al.} implemented a comprehensive IP protection method for GAN models utilizing the regularization terms \cite{ong2021protecting}. This approach visualizes the specific watermarks on the generated images to effectively prevent ambiguity attacks. Yu \emph{et al.} employed image steganography to pre-embed watermarks into the training data \cite{yu2021artificial}. This technique ensures the transferability of the watermarks from the training data to the GAN. Consequently, the watermark encoded in the training data can be decoded from any of the generated images.

    \textit{\ding{173} Watermarks on Latent Diffusion Model.} Pierre \emph{et al.} achieved invisible watermark embedding in the generated images by quickly fine-tuning the latent decoder of the image generator and inserting the binary signature to the generated images \cite{fernandez2023stable}. The hidden watermarks within the generated images are recovered using a pretrained watermark extractor, and statistical tests are employed to determine whether the images are generated from the target model.

    \textit{\ding{174} Watermarks on Large Language Model.} Abdelnabi \emph{et al.} successfully embedded watermarks to natural language models \cite{abdelnabi2021adversarial}. They introduced the adversarial watermarking transformer (AWT) to generate output texts with implicit encoding of a given input text and binary message through collaborative training of the encoder-decoder and adversarial training. Notably, it is highlighted that this scheme is applicable to large language models, such as GPT-2.

    \item \emph{Watermarks on AI-Produced Contents.} For AI-produced content, watermarking methods can be categorized as static watermarks (for hard IP) and dynamic watermarks (for soft IP) based on detection methods \cite{7154454}.

\textit{\ding{172} Static Watermarks.} Static watermarks are inserted when the content is generated by AI and remain constant throughout the content's lifespan. They can be either visible or invisible \cite{hua2016twenty}. Visible watermarks are often used for copyrighted content that companies or individuals wish to market. Invisible watermarks remain hidden within the content. Static watermarks facilitate the tracking of content distribution across various platforms by monitoring unauthorized use or replication of the content. While static watermarks are permanent and relatively easy to implement, they can be more susceptible to removal due to their static nature \cite{mohanarathinam2020digital}.

 \textit{\ding{173} Dynamic Watermarks.} Dynamic watermarks are dynamically generated within the content and can be modified or removed post-generation \cite{7154454}. They are typically used to create a unique and identifiable version of the content each time it is accessed so as to track the content's usage. Dynamic watermarks prevent the misuse of content and facilitate digital forensics, particularly in instances where the content is reused or exchanged on a collaborative platform. While dynamic watermarks are more challenging to remove, they can also complicate the workflow.

\item \emph{Watermarks on AIGC Models and AI-Produced Contents.} Watermarks on both AIGC models and their resulting outputs can be classified into two categories.

\textit{\ding{172} Separated manner}. The watermarks for AIGC models and their produced contents are added in a separated manner. For the content watermark, it pertains to the post-generation addition of content watermarks using conventional multimedia watermarking techniques such as those used for images \cite{mohanarathinam2020digital}, audio \cite{kamaruddin2018review}, text \cite{hua2016twenty}, and others; for the model watermark, it can be embedded through two methods: by incorporating the watermark into the model parameters or by utilizing a trigger set for embedding.

\textit{\ding{173} Coupled manner}. It lies in coupling AIGC model watermarks with generated content. This involves embedding watermarks at the time of content creation using a watermark extractor or adding watermarks within the training data, thereby implementing the watermark-adding process in AI-produced contents. For instance, a watermark extractor is simultaneously trained during the training process of the AIGC model. Combining the loss of the watermark extractor with the loss of the generated model ensures that all the generated content contains extractable watermarks \cite{wu2020watermarking}. It becomes possible to determine the ownership of both the content and the model by extracting the watermark from the generated content. Consequently, the watermarks of content and model are embedded in a coupled manner.
\end{itemize}


\emph{4) Requirements of AIGC Watermarking.} An effective AIGC watermarking scheme should be robust enough to resist model/data transformations and compression, secure enough to defend against sophisticated attacks, faithful enough to maintain the quality of content and performance of the AIGC model, scalable enough to work with different models and contents, and efficient enough to be practical to implement. 
\begin{itemize}
    \item \emph{Robustness.} Effective AIGC watermarks should be robust enough to resist attacks specifically designed to remove the watermark or manipulate the model/content in a way that renders the watermark ineffective. The watermark should be able to withstand various types of attacks, including data tampering, intentional removal, or common distortions.
    \item \emph{Security.} Security is a critical requirement for AIGC watermarking. The watermark should not only be difficult to remove or destroy but should also be resistant to counterfeiting, forging, ambiguity, or overwriting.
    \item \emph{Fidelity.} The watermark should not impact the quality or performance of the AI model. The embedding process should not introduce any significant changes to the model's accuracy and reliability.
    \item \emph{Scalability.} Watermarks should be scalable, indicating that it should be feasible to implement it in various AIGC models without significant difficulty. It should also be possible to embed watermarks on contents of different volumes and dimensions.
    \item \emph{Efficiency.} Watermarks should be efficient in terms of computational overhead and storage requirements. The process of embedding and detecting watermarks should not significantly increase the processing time or storage space needed by the AIGC model. Additionally, the watermarking process should be easy to use.
\end{itemize}

\emph{5) Summary and Lessons Learned.} Existing AI watermarking approaches primarily utilize three approaches: (i) \emph{trigger set} (e.g., backdoors and adversarial examples); (ii) \emph{model parameters} (e.g., trainable parameters or hyper-parameters); and (iii) \emph{unique features}, as the digital fingerprint for ownership verification. When it comes to content watermarking methods, there are two main categories: \emph{static} and \emph{dynamic}.

To ensure comprehensive IP protection and regulation in AIGC, it is usually necessary to embed watermarks on both AIGC models and AI-produced content. Different watermarking techniques can be used depending on the specific requirements and goals of the application, and a combination of watermarks may provide better protection against IP infringement and unauthorized use.
However, due to the constant updating of training data, it is necessary to retrain AIGC models and improve watermarking efficiency. Hence, it is essential to develop an adaptive model watermark embedding mechanism with reduced computational cost during frequent model retraining. Moreover, with the smart and diverse attack behaviors, there is a rising need to enhance the robustness and security of AIGC watermarks. Additionally, it is challenging to implement fast, efficient, and low-cost watermark embedding and verification mechanisms to meet increasing demands for large-scale AIGC data production.

\subsection{Threats and Countermeasures to AIGC Watermarking}\label{subsec:Regu3}

\emph{1) Watermark Distortion.}
Watermark distortion is to distort the results of watermark verification without extra training. In model based watermarks, the adversary can deliberately modify parameters by scaling or repositioning nodes to increase the distance between the embedded vector and the corresponding parameters.
\begin{itemize}
    \item \emph{Permutation Attack.} Permutation attacks \cite{Greedy} aim to permute the nodes of the model, thereby increasing the distance between watermarked parameters and the pre-designed vector. However, permutation attacks can often result in a decrease in model performance, rendering them impractical for real-world applications.

    \item \emph{Scaling Attack.} Scaling attacks \cite{Greedy} aim to scale the model weights by a certain multiple without affecting the output evaluation. In \cite{Greedy}, Liu \emph{et al.} multiply the weights of incoming edges by a certain factor and divide the weights of outgoing edges by the same factor value to preserve the output. Although scaling attacks can distort the distance between the vector and watermarked parameters, the extracted signs (i.e., positive and negative signs) remain basically unchanged, and thus can be used for watermark verification.
\end{itemize}

\emph{2) Watermark Elimination.}
The process of watermark elimination involves creating a replica of the model and removing the embedded watermark through additional training. Adversaries may attempt to remove the watermark in various ways, such as training directly on the host model (i.e., removal attack), compressing the model through knowledge distillation, or even creating a new replica from scratch that duplicates the functionality of the original model (i.e., model extraction).

\begin{itemize}
    \item \emph{Removal Attack.} This attack can be launched through various methods including fine-tuning, pruning, and retraining.

    \textit{\ding{172} Fine tuning.} Fine-tuning is a transfer learning method that involves additional training of the output layer on other tasks, and is often used as an unintentional attack to remove embedded watermarks \cite{white_first}. However, almost all proposed watermarking methods have been shown to be able to resist fine-tuning attacks \cite{passport}\cite{adv_blackbox}\cite{external_features}.

    \textit{\ding{173} Pruning.} Weight pruning is a model compression technique that can also be used to remove embedded watermarks by eliminating redundant nodes in the host model while preserving its performance. Resistance against pruning attacks is a fundamental requirement for watermarking methods, and most works have been shown to be effective against such attacks.

    \textit{\ding{174} Retraining.} Retraining attack is a variation of fine-tuning that aims to remove the embedded watermark by retraining the entire model on a smaller dataset \cite{blackbox_first}, thereby trading off computation cost with the possibility of successful watermark removal. 

    \textit{\ding{175} Data enhancement.} The content-enabled watermarks is to insert special content (e.g., the special customized symbols, company logo) into the output data. The adversaries can remove content-enabled watermarks through data enhancement, i.e., cropping, noise-adding, rotation, mixup.
    \item \emph{Reverse Engineering.} Reverse engineering \cite{neuralcleanse} is a two-stage process that combines detecting hidden watermarks and unlearning to remove the backdoor of a host model. Data-based watermarking methods are vulnerable to reverse engineering threats, as demonstrated by the watermark in \cite{blackbox_first}, which is a typical backdoor watermark with limited robustness against reverse engineering.

    \item \emph{Model Distillation.} Model distillation is a model compression process that involves training a small-scale model, known as the student model, to learn the knowledge of a larger and more complex model, referred to as the teacher model. Theoretically, model distillation can eliminate all model-based watermarks, as the training of the student model is primarily based on the soft and hard labels produced by the teacher model. As such, the parameters of the teacher model can hardly be transferred to the student model.

    \item \emph{Model Extraction.} Similar to model distillation, model extraction aims to replicate the functionality of a remotely deployed model by using black-box queries to retrieve the model's output \cite{extraction,34740853475591}. However, model extraction may result in a larger model and in certain scenarios, can even extract the exact parameters of the host model.
\end{itemize}

\emph{3) Ambiguity Attack.} The ambiguity attack aims to forge an additional watermark to cast doubt on the ownership of a model, without necessarily destroying the formal watermark. During the verification process, an adversary can verify only the additional watermark and consequently claim their ownership. Fan \emph{et al.} \cite{ieee_passport} demonstrate that formal data-based methods (i.e., \cite{white_first}\cite{blackbox_first}) are not robust against ambiguity attacks, and propose a novel passport technique that embeds watermarks in hyperparameters. If the corresponding passport is modified (i.e., re-embed an ambiguous watermark), the model performance would be significantly degraded, which thwarts the original intention of model theft.

\emph{4) Watermark Overwriting.} Once the adversary has knowledge of the watermarking strategy of the owner, they may deploy the same embedding framework to  overwrite the original watermark with a new one. Liu \emph{et al.} prove that the backdoor method \cite{blackbox_first} is not robust against watermark overwriting when the adversary has access to the original dataset. Furthermore, they also demonstrate the robustness of the watermark presented in \cite{Greedy} against watermark overwriting, regardless of whether the adversary can access the original training dataset or not.

{\emph{5) Watermark Collusion.} In federated AIGC settings, several clients may collaborate to train a model and share the model property consistently or independently \cite{fedipr}. Insider attackers may collude to remove, alter, replace, steal, or counterfeit a watermark from the AIGC model/content, and misappropriate the model property with the information obtained through training process. Future research efforts are needed to investigate defense mechanisms against  watermark collusion attacks.}



\section{Future Research Directions}\label{sec:FUTUREWORK}
In this section, we point out several future directions to be investigated in the era of AIGC from the following aspects.

\subsection{Green AIGC Architectures}\label{subsec:future1}
For large AIGC models such as ChatGPT, {they usually require} extensive computing power in both model training and service offering, resulting in significant energy consumption and environmental concerns.
As reported, the energy consumption of ChatGPT in answering a single question for 590 million users is equivalent to the monthly electricity usage of 175,000 Danes \cite{ChatGPTEnergy}. 

Given the current trend towards eco-friendly technologies, the design of green AIGC architecture has become an urgent need to mitigate carbon emissions throughout the entire AIGC lifecycle. This initiative relies on several key technologies, including cooperative cloud-edge computing, hardware innovations, and energy-efficient AIGC algorithms. Particularly, cloud-edge computing \cite{xu2023unleashing,du2023enabling} enables scalable model training and inference with reduced energy costs, where the foundation model deployed on the cloud can process complex data and tasks while the dedicated model deployed on distributed edge devices can handle real-time processing.
The latest hardware innovations, such as neuromorphic chips \cite{schuman2017survey}, can perform specialized computing tasks while consuming less energy compared to traditional processors, which can be integrated into the AIGC architecture to reduce energy consumption. Developing energy-efficient algorithms can also reduce the number of computations required for AIGC models, thereby alleviating the energy demands of model training and inference. Additionally, model compression techniques \cite{deng2020model} such as pruning, quantization, and knowledge distillation can be utilized to reduce the size and complexity of AIGC models, further reducing energy consumption.
Open research problems in this field include optimal cloud-edge resource sharing, automated and adaptive model compression, and dedicated AI chip design, as well as the trade-off between energy consumption and performance.


\subsection{Explainable AIGC Models}\label{subsec:future2}
Current generative AI models, also known as AIGC models, are often considered as black-boxes since they lack transparency and interpretability, making it challenging to understand their decision-making processes. This has raised concerns regarding biased outcomes, security threats, and ethical considerations, resulting in a demand for developing explainable AIGC models.

Several key technologies can be employed in designing explainable AIGC models, including interpretable AI algorithms, model visualization techniques \cite{arrieta2020explainable}, and human-in-the-loop (HITL) approaches \cite{9678840}. Interpretable machine learning algorithms, such as decision trees and rule-based models, enable capturing complex relationships between input features and outputs, thus providing explanations for the model's output. Moreover, model visualization techniques \cite{arrieta2020explainable}, such as activation mapping, saliency maps, and feature visualization, offer a graphical representation of the decision-making process of AIGC models, thus facilitating users' understanding of how models categorize, cluster, or associate different inputs. HITL methods involve the incorporation of human experts in the decision-making process of AIGC models, which can be done through co-designing interfaces and interactive feedback mechanisms for better results \cite{9678840}. The combination of these technologies can improve the transparency and interpretability of AIGC models.

However, several issues remain to be investigated in this area. One of the most significant challenges is to strike a balance between accuracy, interpretability, and security. Black-box models that lack transparency have shown exceptional performance on various data-driven tasks and better security, while interpretable models often exhibit lower accuracy and worse security. Another challenge is the evaluation and validation of explainable AIGC models, requiring a suitable framework to measure and compare their interpretability performance.

\subsection{Distributed \& Scalable AIGC Algorithms}\label{subsec:future3}
With the rising demand for AIGC services and applications across industries, existing centralized AIGC paradigms for model inference and service offering may not be able to handle the expected rise in workload. This will inevitably result in high latency and low scalability in AIGC services. A distributed AIGC paradigm is required, which allows tasks to be mitigated and parallelized, resulting in decreased time required for large-scale AIGC tasks and increased scalability.

Cooperative cloud-edge computing is one of the key technologies that empower distributed and scalable AIGC \cite{xu2023unleashing,9696188}.
It involves the collaboration of cloud and edge devices to distribute computing power and processing tasks across a network of interconnected devices. In the system, a foundation AIGC model is deployed on the cloud to process complex general-purpose tasks, while dedicated AIGC models are deployed on distributed edge devices to handle real-time task processing. The dedicated AIGC models can be optimized through fine-tuning from the larger foundation AIGC or collaboratively trained among a federation of edges, which perform better in specific fields (e.g., medical and transportation sectors). Users requiring services in these sectors can request the smaller-sized dedicated model deployed on edge devices, which results in decreased latency, improved performance, and reduced service costs.

However, several research issues remain to be addressed. Firstly, edge devices may not have sufficient security resources to protect sensitive data. It is necessary to develop effective privacy and security protocols for edge devices to maintain data privacy. Secondly, allocating resources across cloud/edge devices with different capabilities in terms of processing, storage, and communication is challenging. Thirdly, in practical AIGC service architecture, heterogeneous networks can result in a varied quality-of-service (QoS) for different devices connected to the network, while the heterogeneity of these networks complicates the resource management and optimization. Lastly, compatibility issues across edge devices can lead to challenges in load balancing, data transfer, and task allocation.

\subsection{Trustworthy \& Regulatable AIGC Services}\label{subsec:future4}
The rise of AIGC services has sparked concerns about the quality, reliability, and trustworthiness of the generated content. Essentially, AIGC models have the potential to produce disinformation or false content, which can lead to the spread of misinformation, deepfake videos, and propaganda. Moreover, evaluating the authenticity of AI-produced content can be difficult. Additionally, AIGC services should comply with relevant laws and regulations to protect public interest, prevent harm, and ensure accountability.

To ensure the trustworthiness and regulatability of AIGC services, a range of technologies can be integrated including explainable AI \cite{tjoa2020survey}, auditing \cite{asplund2020auditing}, blockchain \cite{9631953,kishigami2015blockchain,lin2020blockchain}, AIGC watermarking \cite{blackbox_first,three_blackbox}, and regulatory compliance frameworks \cite{9268472}. Auditing methods can be employed to eliminate data bias and discrimination in AIGC services while ensuring fairness \cite{asplund2020auditing}. Blockchain technology can enhance trust, transparency and regulatory compliance in AIGC services by recording data transactions, accessing control and user permissions on immutable ledgers \cite{9631953,kishigami2015blockchain,lin2020blockchain}. However, Blockchain raises privacy concerns for data and model collection, storage, and sharing, as well as scalability and performance limitations in practical AIGC services.
AIGC watermarking serves as digital fingerprints and {provides} a means of identifying the ownership of AIGC models and their generated content source \cite{blackbox_first,three_blackbox}. AIGC watermarking can enhance traceability, accountability, and authenticity, while minimizing the risks of model theft and misinformation spreading. However, to meet growing demands for large-scale AI-produced data and alleviate the overhead in frequent model retraining, it requires further investigations to implement watermark embedding and verification mechanisms in an adaptive, fast, efficient, and cost-effective manner.
AIGC models require frequent retraining, adaptive watermark embedding, secure watermarks, and efficient implementation to meet the growing data production demands. Lastly, a combination of hard laws and soft laws is required for regulatable AIGC systems, where hard law regulates the legal aspects of AIGC models and contents while soft law promote ethical practices and standards for their creation, use, and distribution.

\subsection{{Personalized AIGC Services}}\label{subsec:future5}

{AIGC services, such as chatbots and digital assistants, need to interact with massive users with diverse preferences, backgrounds, and needs \cite{du2023enabling}. Therefore, it is essential to provide personalized AIGC services that can adapt to the individual characteristics and preferences of various users, enhancing their satisfaction and engagement. For example, a personalized AIGC chatbot can generate tailored responses according to the user’s profile (e.g., age and gender), behavior, location, and mood.}

{Personalized AIGC services can be built using various key technologies, such as intention modeling and digital twins. Intention modeling focuses on accurately modeling and dynamically updating users' profiles, behaviors, goals, intents, and preferences based on historical multimodal prompts and explicit or implicit feedback. As such, it enables the selection or recommendation of the most suitable AIGC services or responses for individual users. Digital twin technology facilitates the connection between virtual representations and physical entities. It allows AIGC services to create a virtual replica or representation of a real-world object or system, providing personalized and tailored experiences to users \cite{10090432}.}

{However, delivering personalized AIGC services comes with several challenges. One major challenge is the need for substantial amounts of high-quality data to gain a comprehensive understanding of users' characteristics and preferences. The process of collecting such data can be expensive and may raise privacy concerns as it often involves gathering sensitive information. Additionally, the feedback from users can be noisy, inconsistent or adversarial, which may degrade the performance and robustness of AIGC services. Particularly, noisy feedback may lead to inaccurate user profiles or misinterpretation of user preferences, while inconsistent feedback can make it challenging to provide consistent and personalized experiences. Moreover, adversarial feedback may be deliberately misleading or designed to exploit vulnerabilities in the system, further complicating the delivery of personalized AIGC services.}

\subsection{Security-by-Design AIGC Paradigm}\label{subsec:future6}
The increasing use and demand for AIGC services in various applications have made security and privacy concerns more pressing. Developing security-by-design AIGC systems that can resist attacks, protect sensitive data, and preserve user privacy is crucial to mitigate potential threats.
The security-by-design AIGC paradigm ensures that security is not an afterthought, but rather an integral part of the system architecture, making it more robust, resilient, and secure.
Particularly, it takes into account potential security risks associated with AI and embeds security measures into the design and development of AIGC systems. As such, it helps mitigate potential threats to data privacy and security caused by malicious attacks or accidental vulnerabilities in the system. Adversarial robustness \cite{ijcai2021p591} is a key aspect of security-by-design AIGC systems. It refers to the ability of an AIGC system to resist and defend against attacks designed to fool or manipulate it. As such, it ensures that an AIGC system remains reliable and trustworthy, even when faced with unexpected or adversarial inputs. Various techniques, such as robust optimization \cite{shaham2018understanding}, adversarial training \cite{ijcai2021p591}, and defensive distillation \cite{papernot2016distillation}, can improve the adversarial robustness of AIGC systems. Open research problems in the development of security-by-design AIGC systems include the endogenous secure AIGC algorithms, novel proactive defenses, secure coding, rigorous testing, security monitoring, privacy compliance.

\section{Conclusions}\label{sec:CONSLUSION}
The booming of ChatGPT and AI-generated content technologies has led to significant advancements in content creation. However, it has also brought about several security threats such as Jailbreak, data theft, deepfakes, and biased/harmful content.
To keep pace with the advances in AIGC era such as ChatGPT and GPT-4, this survey provides a comprehensive review on AIGC from the security perspective. Particularly, we first introduce the working principles of AIGC including its enabling technologies, general architecture, working modes, key characteristics, and applications. Then, we identify existing/potential security/privacy threats and trust/ethical issues in GPT and AIGC. The state-of-the-art regultion solutions to AIGC are also examized, as well as the challenges and limitations of these solutions. Lastly, essential future directions are discussed for the development of green, explainable, effective, accountable, and secured AIGC frameworks.
The primary objective of this survey is to offer a comprehensive insight into the working principles, security/privacy/ethical concerns, and potential countermeasures in AIGC, alongside motivating further innovative initiatives in the field of AIGC.

\bibliographystyle{IEEETran}

\bibliography{GPTref.bib}

\end{document}